\documentclass{aa}  

\usepackage{graphicx}
\usepackage{txfonts}
\usepackage{lipsum}
\usepackage{subcaption}         
\usepackage{xcolor}             
\usepackage{lscape}             
\usepackage{placeins}           
\usepackage{lineno}
\usepackage{multirow}
\usepackage[colorlinks=true, allcolors=blue, linkcolor=blue, urlcolor=cyan]{hyperref}
\usepackage{soul}
\usepackage{float}
\usepackage{upgreek}
\makeatletter
\def\@fnsymbol#1{\ensuremath{\ifcase#1\or
\dagger\or \ddagger\or \mathsection\or \mathparagraph\or
\|\or **\or \dagger\dagger\or \ddagger\ddagger\else\@ctrerr\fi}}
\makeatother

\begin{document}

   \title{The Origin of Multi-TeV Gamma-rays in  LHAASO J0341+5258 via Cosmic Ray Illumination of  Molecular Clouds}


%
%
%

   \author{Abhijit Roy\inst{1,2}\thanks{AR: abhijit.roy@gssi.it}        
            \and Alan Sunny\inst{3,6}\thanks{AS: alan.sunny@inaf.it}
            \and Martina Cardillo \inst{3}\thanks{MC: martina.cardillo@inaf.it}
            \and Jagdish C. Joshi \inst{4,5}\thanks{JCJ: jagdishcjoshi86@gmail.com}
            \and Ritabrata Sarkar \inst{1,2}\thanks{RS: ritabrata.s@gmail.com}
            }

\authorrunning{Roy et al.}
   \institute{Gran Sasso Science Institute (GSSI), Viale Francesco Crispi 7, 67100, L'Aquila (AQ), Italy
   \and INFN - Laboratori Nazionali del Gran Sasso, via G. Acitelli 22, Assergi (AQ), Italy
   \and INAF - Istituto di Astrofisica e Planetologia Spaziali (IAPS), Via del Fosso del Cavaliere 100, 00133 Roma, Italy
   \and Aryabhatta Research Institute of Observational Sciences (ARIES), Manora peak, Beluwakhan, Uttarakhand 263001, India
   \and Centre for Astro-Particle Physics (CAPP) and Department of Physics, University of Johannesburg, PO Box 524, Auckland Park 2006, South Africa
   \and
   Macroarea di Scienze MM.FF.NN., Università di Roma Tor Vergata,
   Via della Ricerca Scientifica 1, 00133, Rome, Italy
   }

   \date{Received \today}

 
  \abstract
{}
{We investigate the origin of the ultra-high-energy $\gamma$-ray emission detected by the Large High Altitude Air Shower Observatory (LHAASO) from the source LHAASO J0341+5258, which has not yet been associated with any known astrophysical object within the detector’s field of view. 
}
{The observed UHE emission is modelled within two independent frameworks: initially with a time-dependent, source-independent hadronic scenario implemented with the numerical package \textsc{GAMERA}, where particles propagate through the interstellar medium and subsequently interact with the molecular gas observed in the region; and finally with an analytical description of the interaction between the accelerated cosmic-ray population from a supernova remnant (SNR) and the molecular gas around it. The relevant parameter space for the hypothetical past SNR is explored using the observed TeV $\gamma$-rays.
}
{We show that, for physically plausible source-cloud separations and propagation timescales, \textsc{GAMERA} provides a first-order approximation to the spectral modifications induced by particle transport and yields an adequate fit to the ultra-high-energy $\gamma$-ray data. Within the framework of our analytical approach, we demonstrated that the observed GeV emission can plausibly originate from the SNR itself, while the TeV emission detected by LHAASO can be consistently interpreted as arising from particles that have escaped from the SNR and are illuminating nearby molecular clouds.
}
{We invoke a spatio-temporally evolved SNR-molecular cloud interaction scenario to account self-consistently for the entire $\gamma$-ray spectrum from GeV to TeV energies. Despite the remaining uncertainty regarding the nature of the acceleration source, we conclude that the TeV emission detected by LHAASO can be consistently interpreted within the framework of an illumination scenario.
}

   \keywords{Giant Molecular Clouds --
                SNR-GMC interaction --
                LHAASO -- Unidentified sources
               }

   \maketitle
   \nolinenumbers

\section{Introduction}
\label{sec:intro}
The detection of ultra-high-energy (UHE; $E_{\gamma} \geq 100$ TeV) $\gamma$-rays from Galactic sources by the Large High Altitude Air Shower Observatory (LHAASO) has opened a new window on the most extreme particle accelerators in the Milky Way \citep{lhaaso21,lhaaso24}. $Gamma$-rays with energies above $\sim 100$ MeV can be produced through leptonic, such as Bremsstrahlung and inverse Compton scattering, and hadronic interactions, where accelerated protons or nuclei interact with ambient matter ($p$-$p$) or radiation fields (p-$\gamma$) to produce neutral and charged pions that subsequently decay into $\gamma$-rays and neutrinos. A particularly robust signature of hadronic cosmic-ray (CR) interactions is the characteristic ``pion bump’’ observed around 100 MeV, which arises from the rest mass of the $\pi^0$ meson \citep{Yang2018A&A}. On the other hand, the detection of $\gamma$-rays extending beyond 100 TeV implies the presence of particles accelerated to PeV energies, identifying these sources as candidate PeVatrons \citep{cardillo2023AppSc_6433C}. The discovery of photons at such extreme energies has significantly advanced our understanding of Galactic particle acceleration, but it also represents a big challenge. Indeed, it shows that UHE $\gamma$-ray emission alone does not uniquely determine the nature of the underlying particles, since both hadronic interactions and highly energetic leptons, under suitable conditions, can produce photons in this energy range \citep{Breuhaus2021ApJ_9B}. Disentangling these scenarios therefore requires detailed spectral, morphological, and multi-messenger studies.

In the standard paradigm, supernova remnants (SNRs) are considered the primary accelerators of Galactic CRs, a picture supported by the detection of the characteristic pion-decay signature \citep{Giuliani11,Ackermann13}. However, despite strong evidence for hadronic acceleration, no SNR has yet been firmly detected above 100 TeV. This may reflect the fact that the most efficient particle acceleration occurs only during the early stages of SNR evolution \citep{Bell13,Cardillo15,Gaggero2018MN_7G}. Nevertheless, UHE emission may persist long after this PeVatron phase if escaped CRs become trapped and interact with nearby dense molecular clouds (MCs), producing extended $\gamma$-ray emission through hadronic interactions \citep{Mitchell&Celli24}. At the same time, recent observations by LHAASO have revealed very-high-energy (VHE) and UHE $\gamma$-ray emission from a growing variety of source classes, including young massive star clusters (YMSCs), microquasars (MQs), and pulsar wind nebulae (PWNe) \citep{lhaaso24}. These discoveries suggest that the Galactic CR population may arise from multiple accelerator classes rather than from SNRs alone.

Addressing these questions requires observations across the entire MeV-TeV energy range. Ground-based facilities such as High Altitude Water Cherenkov Experiment (HAWC) and LHAASO have already revealed a population of Galactic UHE emitters, while previous-generation imaging Atmospheric Cherenkov telescopes (IACTs), such as High Energy Stereoscopic System (H.E.S.S), Major Atmospheric Gamma Imaging Cherenkov Telescopes (MAGIC), Very Energetic Radiation Imaging Telescope Array System (VERITAS), established their GeV-TeV counterparts. The next generation of observatories, particularly the Cherenkov Telescope Array Observatory (CTAO) and the ASTRI Mini-Array \citep{CTA19, Scuderi22, Vercellone22}, will provide the angular resolution and sensitivity needed to resolve these systems morphologically and spectrally, offering critical insights into the nature of Galactic PeVatrons and their surrounding environments \citep{Bose2022EPJS_3B,cao2023ARNPS_341C}. We have shown in the past that target materials such as MCs play an important role in multimessenger signals, and the Aquila Rift has been found to be spatially correlated with the IceCube Galactic emission \citep{Roy2023JCAP_47R,Roy_2024JCAP_74R}. At the same time, MCs may provide an important clue to understand another intriguing population within the LHAASO catalogue: the unidentified (UNID) and dark PeVatron candidates. Several of these sources lack an obvious counterpart or a clearly identifiable accelerator capable of powering the observed UHE emission. Nevertheless, a number of them exhibit spatial coincidence with dense MC complexes, raising the possibility that the clouds themselves play a key role in shaping the observed $\gamma$-ray emission and preserve the signatures of past extreme particle acceleration in their vicinity.

Indeed, previous works have proposed that $\gamma$-ray emission can arise when CRs accelerated by nearby or even distant SNRs diffuse into and passively illuminate dense MCs \citep{Gabici_MC_Illumination2009,Mitchell&Celli24}. In such a scenario, the clouds effectively serve as long-lived reservoirs of CRs. This makes passive molecular-cloud "illumination" a compelling framework for explaining at least a fraction of the UNID UHE sources detected by LHAASO. 

Among the sources reported by LHAASO \citep{lhaaso21,lhaaso24}, LHAASO J0341+5258 is a particularly intriguing example. The absence of any clearly identified powerful accelerator within its vicinity, combined with the strong spatial association between the observed $\gamma$-ray emission and the surrounding molecular-cloud population, naturally motivates an alternative interpretation. In this context, an important question emerges: how efficiently can MCs retain and sustain the signatures of past PeVatron activity?

\begin{table*}
    \centering
        \renewcommand{\arraystretch}{1.25}
    \setlength{\tabcolsep}{4pt}
    \begin{tabular}{ccccccc}
        \hline\hline
       Cloud&$\ell$   & $b$     & $d$       & $R$      & $M$ ($H_2 + H_I$)  & n($H_2 + H_I$)\\ 
            &  (deg)  &  (deg)  & (kpc)          &  (pc)       &  (M$_{\odot}$)    &  (cm$^{-3}$)          \\ \hline
        A   & 147.0   & -1.74   & $<$ 0.4        & 0.52        & 13  $\pm$ 4         & 1200 $\pm$ 480        \\
        B   & 146.9   & -1.46   & $<$ 0.3        & 0.92        & 34  $\pm$ 4         & 740 $\pm$ 100         \\
        C   & 147.0   & -1.51   & 0.9 $\pm$ 0.4  & 1.6         & 360 $\pm$ 130       & 1700 $\pm$ 640        \\
        D   & 146.6   & -2.06   & 0.5 $\pm$ 0.4  & 0.99        & 45  $\pm$ 26        & 650 $\pm$ 390         \\
        E   & 147.2   & -1.77   & 4 $\pm$ 1      & 1.2         & 400 $\pm$ 88        & 4500 $\pm$ 1100       \\ \hline

Cloud & $\ell$ & $b$ & $d$ & $R^*_{\mathrm{min}}$ & $M(\mathrm{H}_2)$ & $n(\mathrm{H}_2)$ \\ \hline
1a & 147.2 & -1.36 & $3.7 \pm 0.9$ & 2.4 & $150 \pm 83$ & $104 \pm 59$ \\

1b & 147.2 & -1.31 & $1.5 \pm 0.8$ & 0.37 & $9.4 \pm 5.4$ & $1800 \pm 1100$ \\

1c & 147.1 & -1.27 & $1.2 \pm 0.5$ & 0.68 & $38 \pm 22$ & $1200 \pm 660$ \\

1d & 147.2 & -1.37 & $<0.2$ & 0.077 & $0.29 \pm 0.19$ & $6200 \pm 4100$ \\

2 & 146.9 & -1.30 & $0.73 \pm 0.6$ & 1 & $100 \pm 52$ & $960 \pm 500$ \\

3a & 146.6 & -1.30 & $0.87 \pm 0.5$ & 0.39 & $32 \pm 18$ & $5100 \pm 2800$ \\

3b & 146.5 & -1.35 & $1.5 \pm 0.7$ & 0.43 & $45 \pm 20$ & $5400 \pm 2400$ \\

3c & 146.5 & -1.27 & $2.1 \pm 0.8$ & 0.49 & $24 \pm 9$ & $2000 \pm 780$ \\

3d & 146.5 & -1.31 & $1.9 \pm 0.7$ & 0.44 & $18 \pm 7.2$ & $2000 \pm 800$ \\

4 & 146.3 & -1.42 & $1.4 \pm 0.7$ & 0.57 & $110 \pm 48$ & $5500 \pm 2500$ \\

5 & 146.6 & -1.54 & $1.1 \pm 0.7$ & 0.23 & $30 \pm 14$ & $25000 \pm 12000$ \\

6 & 146.2 & -1.70 & $<0.24$ & 0.32 & $8.8 \pm 2.4$ & $2700 \pm 730$ \\

7 & 146.2 & -2.17 & $0.69 \pm 0.4$ & 0.37 & $110 \pm 78$ & $20000 \pm 15000$ \\ \hline

    \end{tabular}
    \caption{At the top, we list the physical properties of five molecular clouds derived by \cite{Tsuji25} from observations of CO and 21-cm continuum emission lines in the vicinity of LHAASO J0341+5258. In the bottom panel, we present the physical properties of the subset of these clouds that lie within the 68\% containment radius of the LHAASO source and are located at a distance of approximately 1 kpc, as reported in \cite{Tsuji25}.}
    \label{tab:MC_tab}
\end{table*}

\begin{figure}
    \centering
    \includegraphics[width=0.91\linewidth]{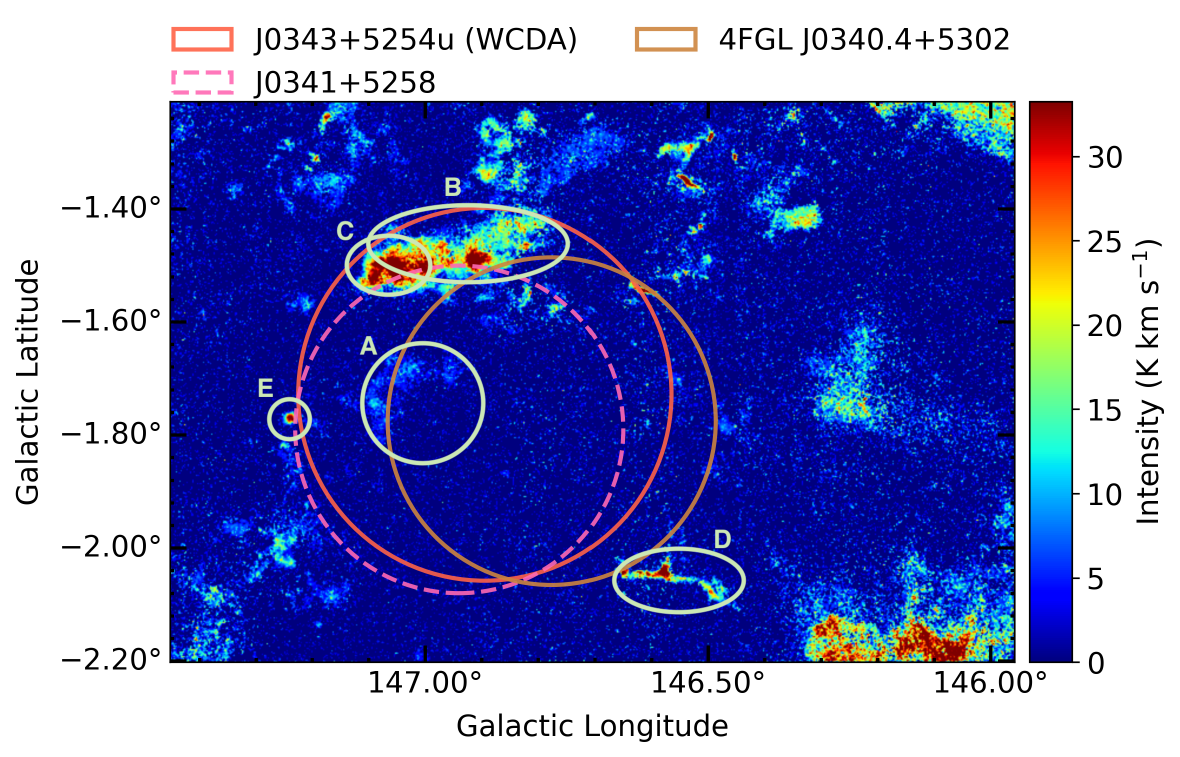}
    \caption{The $^{12}$CO (J=$1-0$) intensity map integrated over the velocity range $V_{\mathrm{LSR}} = -40$ to $10~\mathrm{km~s^{-1}}$, overlaid with the regions observed by the LHAASO and Fermi-LAT instruments toward the very-high-energy $\gamma$-ray source LHAASO J0341+5258, together with the five molecular clouds (shown by white circles), identified within this field of view by \cite{Tsuji25}.}
    \label{fig:roi}
\end{figure}

\subsection{The unidentified source LHAASO J0341+5258}
\label{sec:0341}

In 2019-20, the km$^2$ array (KM2A) detector of the LHAASO observatory detected a new unidentified UHE $\gamma$-ray source, named LHAASO J0341+5258, at R.A.\,$= 55.34^\circ \pm 0.11^\circ$ and decl.\,$= 52.97^\circ \pm 0.07^\circ$ with an exposure of $\sim 98$ days \citep{lhaaso21}. Using the HAWC interactive online tool, they reported a flux upper limit of $1.1\times10^{-12}$ $ \rm erg\;cm^{-2}s^{-1}$, considering a spectral index of $-2.5$, a source extension of 0.5$^\circ$, and a reference energy of 7~TeV with significance of $2\sigma$. Its extension, together with its UHE $\gamma$-ray emission with a hard spectrum makes it a perfect candidate Pevatron. They found no significant detection of very high energy (VHE, $0.1 \lesssim E_{\gamma} \lesssim 100$ TeV) or UHE $\gamma$-rays from the region from other observatories. \cite{Bangale23} also studied the spatial association of this region using VERITAS and HAWC. In their study, they did not find a significant signal in VERITAS, deriving only upper limits; however, they reported HAWC detection with a significance of $\sim8.4\sigma$. In the low-energy $\gamma$-rays, the Fermi 4FGL J0340.4+5302 is the only unidentified GeV $\gamma$-ray source within the angular extension of the LHAASO J0341+5258 with no known companion in any other wavelength \citep{lhaaso21, sarkar24}. Its point-like nature and the curved spectrum with a sharp cutoff at around $\sim$ 2 GeV indicate that this source could be a pulsar \citep{lhaaso21, sarkar24}, but in several attempts, no radio pulsations were detected from this source \citep{Smith2023ApJ_91S}. The absence of any detected pulsation or periodicity leaves its true nature uncertain. In addition, the $\gamma$-rays of Fermi 4FGL J0340.4+5302, could be produced via CR interactions with the dense gas clumps, whose nature we will describe in the later part of the paper. The upper limit set by the Fermi-LAT observation above a few GeVs indicates that the spectrum of LHAASO J0341+5258 at this energy could be relatively hard with a spectral index less than 2. 

The region located at an angular separation of 0.6$^\circ$ from the center of LHAASO J0341+5258 is spatially coincident with several X-ray point sources detected by XMM-Newton \citep{DiKerby25}. These sources are also listed in the second ROentgen SATellite (ROSAT) all-sky survey source catalog and lie within the 0.1-2.4 keV energy band. Owing to the limitations of the ROSAT all-sky survey, the fluxes of these sources are not accurately constrained. Nevertheless, the positions of two of them (2RXS J033928.5+530720 and 2RXS J034316.5+524331) are consistent with those of two sources in the 2SXPS Swift X-ray Telescope point-source catalog, namely 2SXPS 172133 and 2SXPS 171354 \citep{Evans20}. More recently, eleven further X-ray point sources and one extended X-ray emission region (XMMU 034124.2+525720) within the spatial extension of the LHAASO source were identified using data from the XMM-Newton Science Archive \citep{DiKerby25}. Among these eleven point sources, only two (sources ID 7 and 8 in \citet[Table 2]{DiKerby25}) have identified counterparts, specifically a ROSAT X-ray point source and the star UCAC4 715-027962 with apparent magnitude $V = 12.5$.

During February–March 2024, \cite{Tsuji25} conducted observations of the LHAASO J0341+5258 region using the 45-m radio telescope at the Nobeyama Radio Observatory (NRO), with a total integration time of 36 hours. The telescope operates in millimetre waves to observe the emission lines from CO to determine the properties of molecular hydrogen ($H_2$) in the region. For the atomic hydrogen ($H_I$), they use the 21-cm continuum emission line from the Canadian Galactic Plane Survey (CGPS) archival database. The observation of continuum line emission at 408 and 1420 MHz was done from 2003 to 2004 using Dominion Radio Astrophysical Observatory (DRAO) \citep{CGPS}. Combining both the data, they calculate the total column density in the region and the relative number density, assuming a spherical symmetry of the clouds. Five compact clouds of radius less than 2 pc and at a distance $\le$ 4 kpc are detected within the LHAASO UHE source area (see Fig.~\ref{fig:roi}). The properties of these five clouds are tabulated in the first part of Tab \ref{tab:MC_tab}. These clouds of compact size and high density of hydrogen could provide enough target material for any accelerated hadronic cosmic particles to interact and produce $\gamma$-ray or neutrinos from $p$-$p$ interactions. So, the observation of $\gamma$-ray or neutrinos from them can reveal the presence of any cosmic accelerator around them. It is important to note that these five MCs are identified within the 39\% containment radius of the emission region. However, additional MCs located beyond the immediate J0341 extension may also contribute to the observed $\gamma$-ray emission. These clouds were identified by \cite{Tsuji25} and are listed in the second part of the Table~\ref{tab:MC_tab}. Their contribution becomes particularly relevant when considering the two-dimensional Gaussian morphology with a 68\% containment radius, given that the source itself exhibits an extension of approximately $\sim 1^\circ$ \citep{Tsuji25}. The presence of multiple dense gas structures across the region therefore prompts us to explore the possibility that part of the observed UHE emission originates from interactions between accelerated particles and the molecular environment. This scenario is particularly intriguing for a source lacking a clearly identified accelerator, motivating a detailed investigation of the role that MCs may play in shaping the observed $\gamma$-ray and neutrino emission.

The paper is organized as follows. In Sect. \ref{sec:emission}, we introduce the particle transport scenario and emission mechanisms for our source. In subsection \ref{Sec:ProtonDist}, we first evaluate the contribution of the diffuse Galactic CR population to the MCs in the region; we then investigate a source-independent illumination scenario in which particles injected by a distant accelerator diffuse into the clouds, using the numerical package \textsc{GAMERA} in Sec.~\ref{Sec:snr}. Motivated by the resulting energetics and a possible SNR interpretation, we subsequently construct a self-consistent model of the GeV-TeV emission that combines contributions from the Galactic CR sea, SNR acceleration, and SNR illumination of nearby MCs in Sec. \ref{Sec:AccProtonDist}. The corresponding neutrino emission is presented in Sec. \ref{Sec:neutrinos}. Finally, we summarize our results and conclusions in Sec. \ref{Sec:conclusion}.

\section{Modelling the Emissions of Radiations from the Source Region}
\label{sec:emission}

Recently, \cite{kar22} explored a past-explosion model for LHAASO J0341+5258, proposing that an energetic event occurred at the source’s current location approximately 2,000 years ago. Their findings suggest that a brief injection period, lasting about one year, could have successfully populated the MCs with confined particles, effectively reproducing the observed TeV emission. However, the lack of direct observational evidence for such an explosion remains a significant caveat. Subsequently, \cite{sarkar24} modeled the broadband SED of the LHAASO source by invoking a combined lepto-hadronic contribution from an SNR+MC interaction and with a TeV halo scenario. In both cases, the GeV emission is attributed to synchro-curvature radiation originating from a putative pulsar. Motivated by the recent detection of five MCs in the field of view of LHAASO J0341+5258 \citep{Tsuji25}, as well as a point-like X-ray source in the same region \citep{DiKerby25}, we revisit this problem. But because no pulsed emission was detected from the region, we focused on the hadronic emission model instead. 

Compared to previous studies, here we investigate the contribution of CRs originating from non-local sources. These contributions may arise either from the diffuse Galactic CR background or from particles that escaped accelerators located offset from the MC region. The concept of MC illumination by external CR populations was first proposed by \citet{Aharonian_1996emissivity} and later explored in detail by \citet{Gabici_MC_Illumination2009}. More recently, this passive illumination scenario has been revisited in the context of UHE $\gamma$-ray sources by \citet{2022PASJ...74..625O, Mitchell_Illumination2024lhaaso,Mitchell&Celli24, Oka_2025ApJ...989..137O}. In this work, we investigate the origin of the UHE $\gamma$-ray emission within a hadronic $p$-$p$ interaction framework. We first estimate the contribution of diffuse Galactic CRs to the MC illumination scenario, and then examine whether the observed UHE emission can be reproduced through illumination from escaped CRs from an accelerator located away from the clouds, without explicitly assuming a specific source of origin. Motivated by the resulting energetics and transport properties, which point toward a possible SNR association, we further explore a scenario in which a distant SNR together with the Galactic diffused CR population consistently explains the observed GeV-TeV emission.


\subsection{Galactic Diffused Cosmic-Ray Interactions with Molecular Clouds}
\label{Sec:ProtonDist}
\begin{figure*}
    \centering
    \includegraphics[height=0.37\linewidth, width=0.48\linewidth]{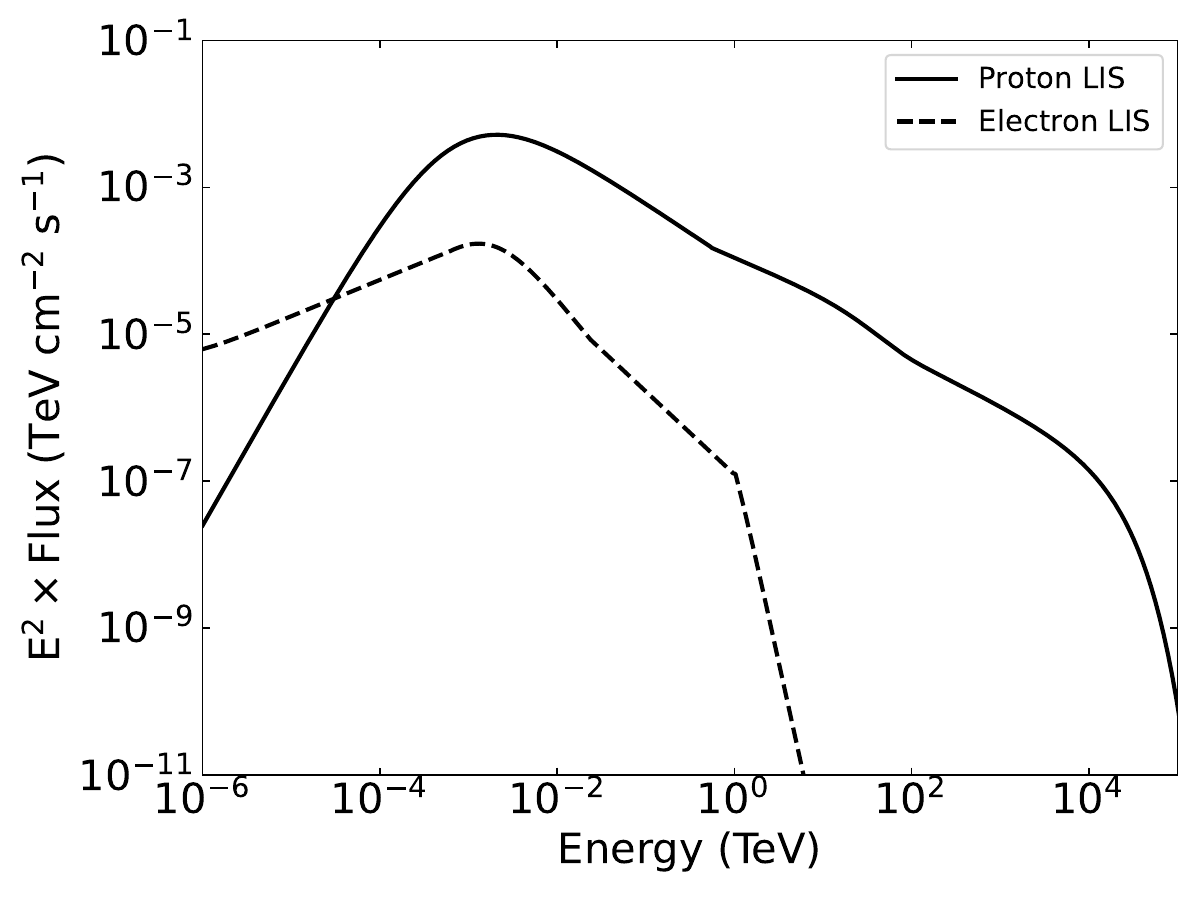}
    \includegraphics[height=0.37\linewidth, width=0.48\linewidth]{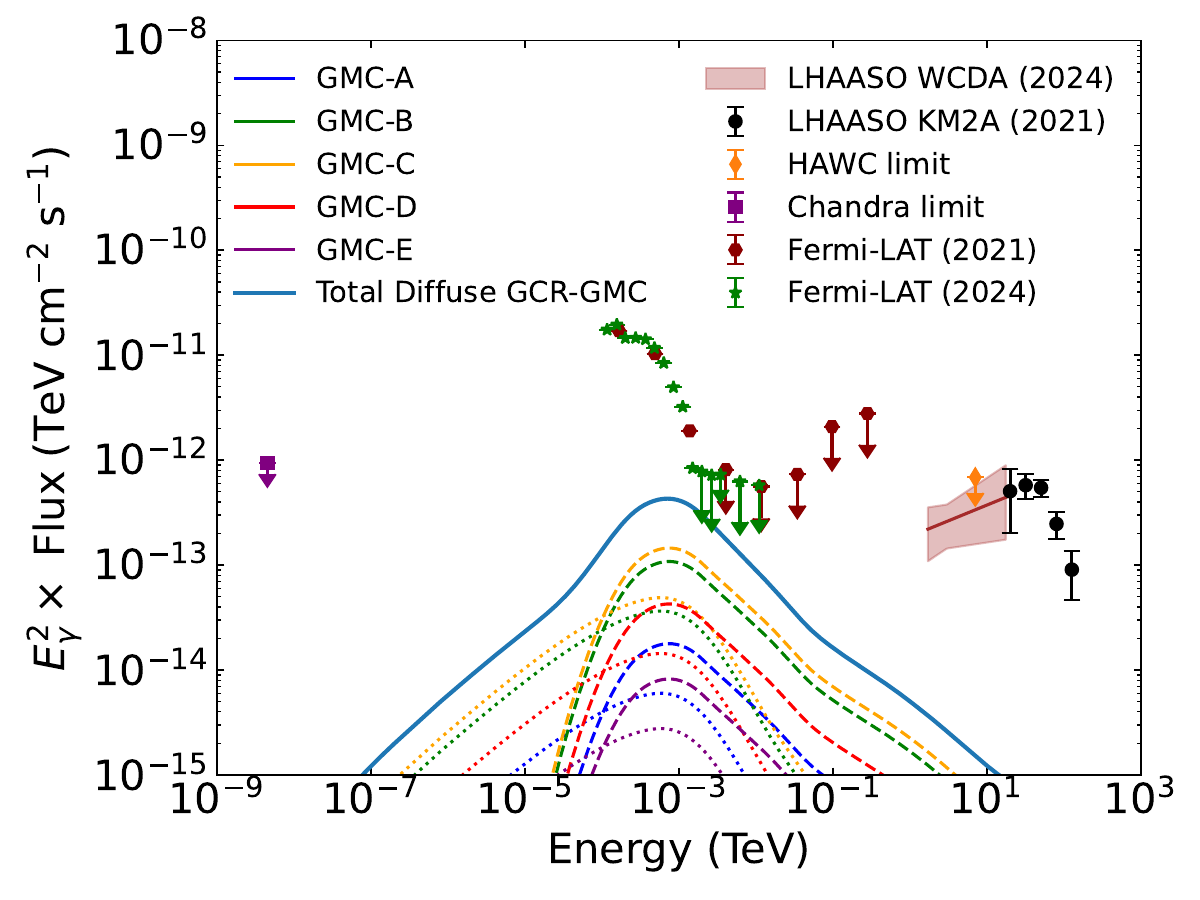}
    \caption{{\it Left panel}: The distribution of primary GCR protons and electrons in the local interstellar medium. {\it Right panel}: $\gamma$-ray emission originating from five GMCs interacting with the diffuse GCR proton and electron populations is shown. The dashed curves represent the resulting $\gamma$-ray spectra produced in each individual cloud via $p$–$p$ interactions, whereas the bremsstrahlung emission associated with the LIS of electrons is depicted by the dotted curves in the same figure.}
    \label{fig:lis_clouds_side_by_side}
\end{figure*}

The Galactic diffused CR component, primarily proton and heavy nuclei present in the region, may interact with the available matter present there and produce HE $\gamma$-rays and neutrinos through hadronic $p$-$p$ interactions. Similarly, the diffuse electrons may also produce HE $\gamma$-rays from leptonic processes such as inverse Compton interaction with the Cosmic Microwave Background (CMB), interstellar radiation field (ISRF) of our Galaxy and synchrotron emission in the intergalactic magnetic field. In order to calculate their contribution, here we consider the spectral distribution of GCR protons and electrons observed near the Earth. The Local Interstellar Spectrum (LIS) for protons can be approximated using the spectral shape described in \citet[Eq. 3.1]{Roy_2024JCAP_74R} (see Fig. \ref{fig:lis_clouds_side_by_side}, black solid line left panel) \citep[originally from]{2023PhRvL.131o1001C, 2015ApJ...815..119V}. The distribution of electrons is approximated by the spectral shape outlined in \citet[Eq. 1, 2]{Potgieter13} for energies below 1 TeV, and by the shape defined in \citet[Eq. 1]{hess24} for energies above 1 TeV (see Fig. \ref{fig:lis_clouds_side_by_side}, black dashed line in left panel). 

We first irradiate the five newly identified MCs with the diffuse GCR proton LIS and compute their individual, as well as cumulative, contributions to the resulting $\gamma$-ray emission. For this purpose, we employ the same formalism as in \cite{Roy_2024JCAP_74R} for evaluating the $\gamma$-ray and neutrino fluxes from a GMC of mass \textit{M} located at a distance \textit{d}. The production of $\gamma$-rays via $p$-$p$ interactions is calculated using the parametrization of the differential cross-sections provided in \cite{Kafexhiu14}, adopting the \textit{SIBYLL} interaction model. To account for the presence of heavy nuclei in both the GCRs and the GMCs, we implement a Nuclear Enhancement Factor (NEF) of 2.09 \citep{Kachel_2014ApJ_36K}. The resulting $\gamma$-ray spectra from the individual clouds are shown in the right panel of Fig. \ref{fig:lis_clouds_side_by_side} as dashed curves. The electron-induced contribution, dominated by bremsstrahlung emission, is plotted as dotted curves in the same figure. Comparison of these model predictions with the $\gamma$-ray fluxes measured by Fermi-LAT and LHAASO indicates that, if the emission from these clouds arises exclusively from interactions between GCRs and GMCs, the resulting flux will be well below the observed levels at both GeV and TeV energies. This discrepancy implies that additional emission components or alternative particle populations are required to account for the observed data.

\subsection{Cosmic rays from distant accelerators: source independent modelling}
\label{Sec:snr}

As introduced above, a compelling narrative for unidentified LHAASO sources could be the "distant accelerator" scenario, where the CR source is located at a spatial offset from the target MC \citep{Aharonian_1996emissivity,Gabici_MC_Illumination2009}. \citet{Gabici_MC_Illumination2009} argued that an SNR positioned at a sufficient distance ($\sim 50$ pc) could effectively illuminate a cloud through escaping CRs while remaining undetected by IACTs. Later, similar illumination models were explored extensively for different sources \citep{Mitchell_Illumination2024lhaaso, Mitchell&Celli24}. For LHAASO J2108+5157, \citet{Mitchell_Illumination2024lhaaso} found that a young SNR ($< 10$ kyr) located within 40-60 pc of the cloud could potentially explain the emission. However, considering that different regions of the Galaxy may host other accelerators, it is also plausible that particles escaping from any accelerator environments can eventually be injected into dense MC structures. So, in contrast to most studies, it may not be just SNRs "illuminating" MCs. This framework is further motivated by recent studies of MQs, in which CRs injected during a past active phase remain trapped within the surrounding cocoon and subsequently generate bright $\gamma$-ray emission through $p$-$p$ interactions with nearby MCs, long after the central engine has become inactive \citep{Abaroa2026_MQRasPeVatrons}.  So we begin by considering a source-independent modelling. Crucially, as the physical distance between the accelerator and MC increases, the particles inevitably undergo energy-dependent transport effects. Consequently, the spectrum of the CRs reaching the cloud is highly sensitive to the source-cloud separation and the diffusion properties of the ISM that are central to the model presented here. 

Here, we adopt a source-independent injection spectrum and use the particle population that reproduces the observed emission to infer the nature of the underlying accelerator. The spectral modelling for the source-independent scenario is performed using \textsc{GAMERA} \footnote{\url{https://github.com/libgamera/GAMERA}} \citep{GAMERA_Hahn2022}, an open-source library designed to simulate the time-dependent evolution of non-thermal leptonic and hadronic populations. While previous studies, such as those by \citet{kar22} and \citet{sarkar24}, utilized this framework to model MC illumination, their approaches typically assumed an injection occurring within the cloud itself by a nearby supernova or a remnant. In contrast, we expand from their studies and explore a scenario where the accelerator is located at a spatial offset from the target MC. Given a representative ISM density of 0.1 cm$^{-3}$ and a magnetic field of 2 $\mu G$ \citep{Heerikhuisen_ISM2011, ferriere_2009interstellar, rand1989local}, we demonstrate that for a reasonable source-cloud distance and propagation timescales, \textsc{GAMERA} provides a first-order approximation for the spectral modification induced by propagation. 

To model the non-thermal emission from the cloud, we utilize \textsc{GAMERA} to numerically solve the time-dependent transport equation:
\begin{equation}\label{Eq:Transport}
\frac{\partial N}{\partial t} = Q(E, t) - \frac{\partial}{\partial E} \left[ b(E,t) N(E,t) \right] - \frac{N(E,t)}{t_{D}},
\end{equation}
where $N(E,t)$ is the particle distribution, $Q(E,t)$ represents the source injection term, and $b(E,t)$ accounts for energy loss rate of these particles. The term $t_D$ represents the characteristic diffusion timescale, which accounts for the energy-dependent escape of particles from the system. Following the framework established by \citet{ Aharonian_1996emissivity, kar22}, we define this timescale as: 
\begin{equation}\label{eq:Diff_time}
    t_D = \frac{R_D^2}{4\,D(E)},
\end{equation}
where $R_D$ is the characteristic distance the particles have to diffuse and $D(E) = D_0\left(E/E_0\right)^{\delta}$ is the diffusion coefficient as a function of energy, where $\delta$ = 0.33 assuming a Kolmogorov-type diffusion with $E_0 = 4\,\text{GeV}$ \citep{kar22,Strong04_Diffuse} and $D_0$ is the diffusion coefficient constant.  

Our numerical approach effectively partitions the simulation into two distinct stages: ISM propagation and MC illumination. We approximate the ISM as a uniform region, and within this ISM, we assume the particles traverse a characteristic length scale of 50 pc. Therefore, in the ISM case, $R_D = 50$ pc represents the assumed separation between the accelerator and the cloud, whereas in the MC case, $R_D$ corresponds to the physical size of the cloud. However, while \textsc{GAMERA} accounts for continuous energy losses and temporal evolution, its standard single-zone implementation does not include a spatially explicit (radial-dependent) diffusion. To accurately model the population of escaped particles that traverse the ISM and subsequently inject into the MC, we perform the spatial transport independently within our modelling framework.

\subsubsection{Spatial Propagation in the ISM}
\label{subsec:ISM_prop}

We model the initial CR population, injected into the localized ISM, using a standard power-law spectrum. Within this initial propagation phase, particles undergo continuous energy-dependent cooling, which is treated numerically using \textsc{GAMERA} over an interstellar propagation timescale, denoted by $t_{\text{prop}}$. For a given injected particle population, the spatial probability density after a time $t_{\text{prop}}$ can be described by the Green’s function $G(R_D,E,t_{\text{prop}})$ \citep{Aharonian_1996emissivity,Gabici_MC_Illumination2009}. The corresponding diffusion radius for $t_{\text{prop}}$ is given by $R_{\text{prop}} = 2\sqrt{D(E)t_{\text{prop}}}$ (from Eq. \ref{eq:Diff_time}), giving us the standard three-dimensional Gaussian Green’s function
\begin{equation}\label{eq:greensfunc}
G(R_D, E, t_{\text{prop}}) = \frac{1}{\left(4\pi D(E) t_{\text{prop}}\right)^{3/2}} \mathrm{exp}\left(- \frac{R_D^2}{4D(E)t_{\text{prop}}}\right)
\end{equation}
We integrate this density distribution over the target MC's volume ($V_{\text{MC}}$) to obtain the total particle population entering the cloud complex. The absolute number of arriving particles per unit energy ($\text{erg}^{-1}$) can be evaluated at the boundary as:
\begin{equation}\label{eq:PostISM_MCInj}
\left. \frac{dN}{dE} \right\rvert_{\textrm{MC, inj}} = N_{\text{total}} (E) \times G(R_D, E, t_{\text{prop}}) \times V_{\text{MC}}
\end{equation}
where $N_{\text{total}}(E)$ is the total CR spectrum, after accounting for energy losses, calculated numerically with \textsc{GAMERA}.
A computational challenge encountered when extracting this escaped particle distribution is the appearance of an unphysical "numerical baseline" or fake tail at low energies due to the precision limits of the exponential term in Eq. \ref{eq:greensfunc}. 
To resolve this numerical limitation and restore physical reasoning, we implement a structural horizon cutoff mask based on the relation between the propagation age and the characteristic diffusion timescale across the separation (Eq. \ref{eq:Diff_time}). At a given $t_\text{prop}$,, the exponential term in the Green's function can be written as an energy-dependent spatial suppression factor, $\exp(-t_D /t_{\text{prop}})$, which maps the spatial envelope of the expanding CR shell \citep{Aharonian_1996emissivity}.
High-energy, fast-moving particles, where $t_D \le t_{\text{prop}}$, are actively flooding into the cloud, so they are kept. Conversely, low-energy particles, where $t_D > t_{\text{prop}}$, are moving too slowly to have reached the cloud within the age of the system. Governed by the exponential term, their physical arrival probability is essentially zero, and they are discarded so the exponential tail naturally should drop to absolute zero. 
We therefore enforce a hard numerical cutoff at this boundary ($t_D > t_{\text{prop}} \implies dN/dE = 0$), which cleanly truncates any floating point noise caused by the exponential term while perfectly preserving the natural, smooth physical slope of the arriving particle wave. This cleaned boundary spectrum is used as the input injection spectrum for the final MC zone. Supported by theoretical models of CRs near energetic accelerators \citep{Oka_2025ApJ...989..137O, Gabici_MC_Illumination2009, Schroer2022_CRBubbleMMHD}, the ISM environments surrounding active sources often exhibit slow diffusion characteristics relative to the Galactic average ($D_0 \sim 10^{28}\text{ cm}^2\text{ s}^{-1}$ at $E_0= 4$ GeV \citep{kar22}), which gradually recovers at larger distances ($\sim 50\text{-}60$ pc) \citep{De2024_origin,Schroer2022_CRBubbleMMHD}. To reflect this localized environment, we adopt a slow diffusion normalization of $D_0 \sim 5 \times 10^{26}\text{ cm}^2\text{ s}^{-1}$ for the propagation of particles to the clouds.

\subsubsection{Injection and illumination of the clouds}
\label{subsec:MCillumination}
In the case of the MC, we require the particles to be well confined within the clouds to argue the observed TeV $\gamma$-ray emission. This requires $t_D \geq t_{age}$, where $t_{age}$ refers to the duration of the particle evolution within the MC environment. Within the MC, however, we implement a suppressed diffusion coefficient to ensure the high-density environment remains a better confinement site for the particle population.

Finally, regarding the target morphology, recent observations have revealed multiple MCs within the LHAASO region of interest (ROI). However, the current angular resolution of UHE $\gamma$-ray observations does not allow the individual contributions from these clouds to be clearly separated. Modelling each cloud independently would therefore introduce a highly degenerate and computationally demanding parameter space. Instead, we adopt a simplified approach in which the MC environment is treated as a single representative cloud with averaged physical properties, such as density and magnetic field strength. This provides a computationally efficient yet physically reasonable description of the large-scale interaction scenario and the resulting integrated emission. Although several of the identified MC clumps are locally very dense, their physical sizes are relatively small. As a result, the effective density averaged over the entire emission region is considerably lower. In our model, we adopt a $\gamma$-ray extension of approximately $\sim 1^\circ$, corresponding to the 68\% containment radius derived from a two-dimensional Gaussian model \citep{Tsuji25}. This larger extension requires the inclusion of all MCs at the $\sim 1$ kpc distance listed in Tables \ref{tab:MC_tab}, and the total mass becomes $\sim 10^3 M_\odot$, consistent with previous estimates \citep{kar22} considering a cloud size of $1.67 \times 10^{19}$ cm. However, whereas \citet{kar22} assumed an average gas density of $50~\mathrm{cm}^{-3}$, our modelling favors a lower mean density of $30~\mathrm{cm}^{-3}$, which yields the best agreement with the observed spectral normalization. For the magnetic environment, we assume an average field strength of $\sim 6 \, \mu \mathrm{G}$, consistent with typical values expected in MC environments \citep{Crutcher_2012MagField}. The final parameters adopted for the LHAASO J0341+5258 MC and the surrounding ISM cloud, which provide the best fit to the observed spectrum, are summarized in Table~\ref{tab:gamera_cloud_params}.

\begin{table}
\centering
    \renewcommand{\arraystretch}{1.25}
    \setlength{\tabcolsep}{4pt}
\caption{Parameters set for the ISM and MC for the simulation with \textsc{GAMERA} }
\label{tab:gamera_cloud_params}
\begin{tabular}{ccc}
\hline\hline
\textbf{Parameters} & \textbf{ISM} & \textbf{MC} \\
\hline
Density ($\mathrm{cm}^{-3}$)& 0.1 & 30 \\
Magnetic field ($\mu G$)& 2 & 6 \\
Characteristic distance $R_D  \,(\mathrm{cm})$ & $1.543 \times 10^{20}$ & $1.67 \times 10^{19}$ \\
Constant $D_0$ ($\text{ cm}^2\text{ s}^{-1}$) & $5 \times10^{26}$ & $10^{26}$\\
\hline
\end{tabular}
\end{table}

\begin{figure}
    \centering
    \includegraphics[width=0.91\linewidth]{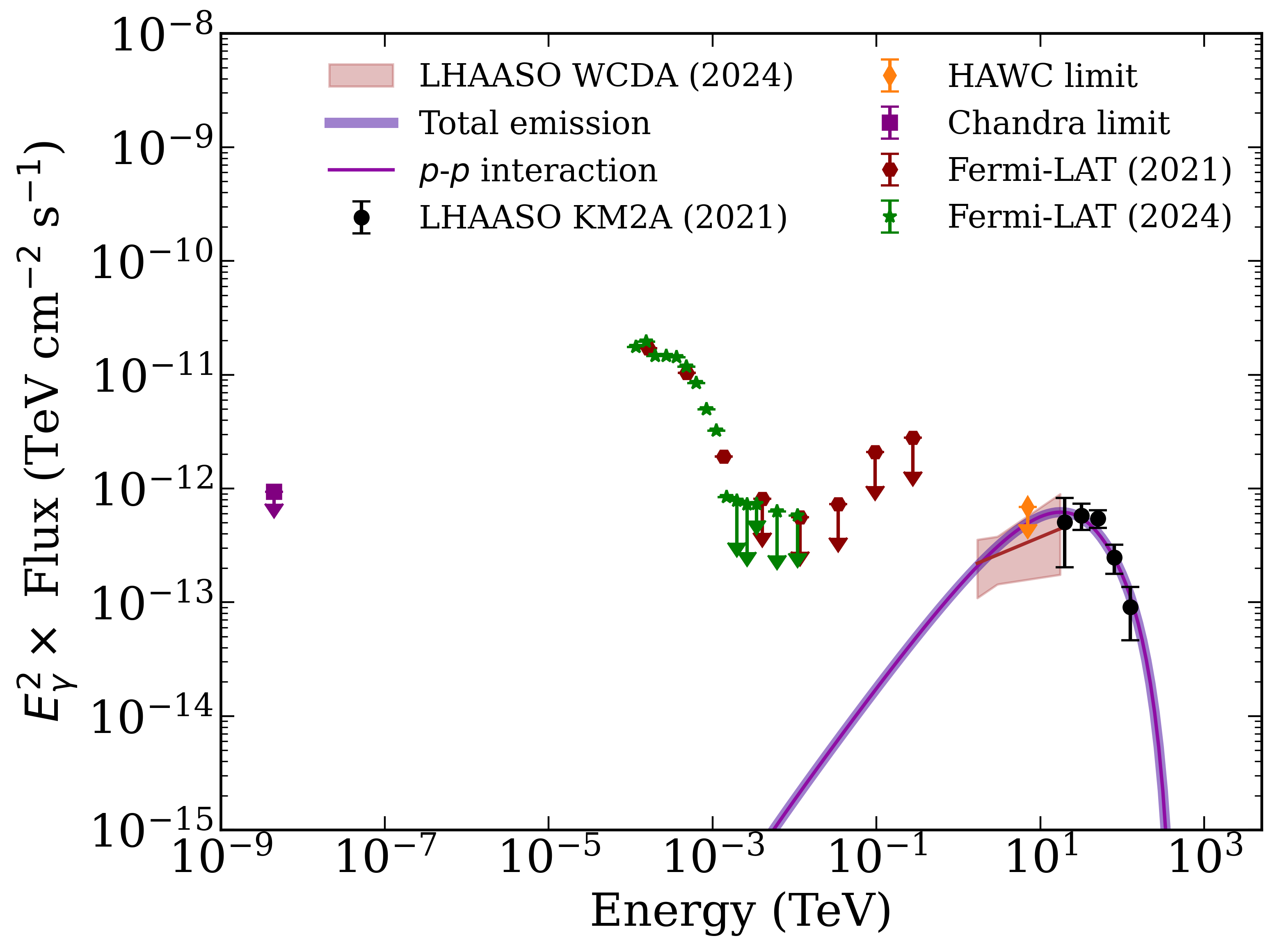}
    \caption{Spectral energy distribution fit using the source-independent illumination scenario, including multiwavelength observations of the emission region. Here we explain the UHE emission modelled via hadronic interactions originating from pion decay (thick solid line). A nuclear enhancement factor of 2.09 is applied to account for the contributions of heavier nuclei. (See Sec.\ref{subsec:MCillumination} for the model details)}
    \label{fig:SED_illum}
\end{figure}

Finally, we present a fiducial set of particle injection parameters chosen to reproduce the spectrum shown in Fig.~\ref{fig:SED_illum}. Although our approach is source-independent, we assume that the underlying accelerator is predominantly hadronic in nature. Under this assumption, particle acceleration is expected to proceed through diffusive shock acceleration (DSA) or non-linear DSA processes. In such environments, high-energy electrons experience severe radiative losses through synchrotron and IC emission, strongly limiting their survival \citep{Aharonian04_Book}. In contrast, relativistic protons can escape the accelerator more efficiently and diffuse into the surrounding ISM and dense molecular environment. Furthermore, the resulting spectral energy distribution will be dominated by $\pi^0$-decay emission from $p$-$p$ interactions, with no significant leptonic contribution required to reproduce the observed $\gamma$-ray spectrum. Motivated by these considerations, we neglect the electron component in the present model and focus exclusively on the hadronic population \citep{Gabici_MC_Illumination2009, Mitchell_Illumination2024lhaaso}. The spectrum in Fig. \ref{fig:SED_illum} was best reproduced considering protons injected into the ISM over a period of $t_{inj} = 1000$ years with luminosity of $8 \times 10^{39}$ erg s$^{-1}$. The injected particle population spans energies from $10^{-1}$ to $600$ TeV for protons with a power-law spectrum of index 1.9. These particles propagate through the ISM for a timescale of $t_{\mathrm{prop}} = 15$ kyr, after which the escaped particle spectrum reaching the MC is extracted following the methodology described in Sec~\ref{subsec:ISM_prop}. The comparison between the injected and the extracted particle population is shown in Appendix \ref{App:Proton_part_story}. The final particle population is then injected and evolved within the extended J0341+5258 MC for $t_{age} =100$ years, producing the final SED shown in Fig.~\ref{fig:SED_illum}. This SED is dominated by $\pi^0$-decay emission resulting from $p$-$p$ interactions. Formally, the total hadronic $\gamma$-ray flux is expressed as $F_{\gamma} = \epsilon_N\, F_{pp}$ where $F_{pp}$ represents the pure $p$-$p$ emission component, and $\epsilon_{N}$ = 2.09 is the nuclear enhancement factor. This factor is applied to incorporate the contribution of heavier nuclei (e.g., He, CNO, MgSi, Fe, etc) in the CRs and the ambient gas \citep{Mori2009_NuclearEnhancement, Kafexhiu14}. 

The parameters that provide the closest agreement with the observed spectrum (Fig. \ref{fig:SED_illum}) offer valuable insights into the nature of the underlying accelerator. The inferred particle energetics and luminosity point to an exceptionally powerful source capable of accelerating particles to several hundred TeV. Moreover, the required particle injection occurs over a relatively short timescale ($t_{\mathrm{inj}} \sim 1000$ yr), suggesting a brief but highly efficient acceleration episode rather than continuous activity over an extended period. Such behaviour is naturally expected during the early evolutionary stages of an SNR, when shock power is at its peak and magnetic field amplification can significantly enhance particle acceleration \citep{Bell2004_turbulent, Ptuskin2005_Spectrum}. Assuming a canonical supernova explosion energy of $10^{51}$ erg, the inferred particle energetics correspond to an acceleration efficiency of approximately 25\%, well within the range often invoked for efficient CR production in SNRs. Furthermore, the required injection spectrum, characterized by a power-law index of $\sim 1.9$, is consistent with the hard spectra predicted by non-linear DSA models of efficient SNR shocks \citep{Caprioli2012_SNR_NDSA}. The combination of the inferred energetics, acceleration timescale, efficiency, and spectral shape collectively points to CRs accelerated in the past by an efficient SNR as a particularly compelling candidate for the origin of the particles responsible for the observed emission.

It is to be noted that the illumination framework does not account for the GeV emission or the reported \textit{Chandra} upper limits. This suggests that the GeV and TeV components may arise from distinct physical regions, with the former can be assumed to be from the accelerator itself or a completely different source and the latter originating from particles confined within the MCs. In the coming section we consider a hypothetical SNR as a plausible accelerator. Since the evolution of an expanding SNR shock introduces spatially and temporally varying transport conditions that cannot be adequately reproduced by simple ISM approximations, and because \textsc{GAMERA} does not permit a fully radial-dependent treatment of particle transport, we adopt an analytical approach to explore this possibility.

\subsection{The SNR hypothesis and interactions with Molecular Clouds}
\label{Sec:AccProtonDist}

\begin{figure}
    \centering
    \includegraphics[width=0.91\linewidth]{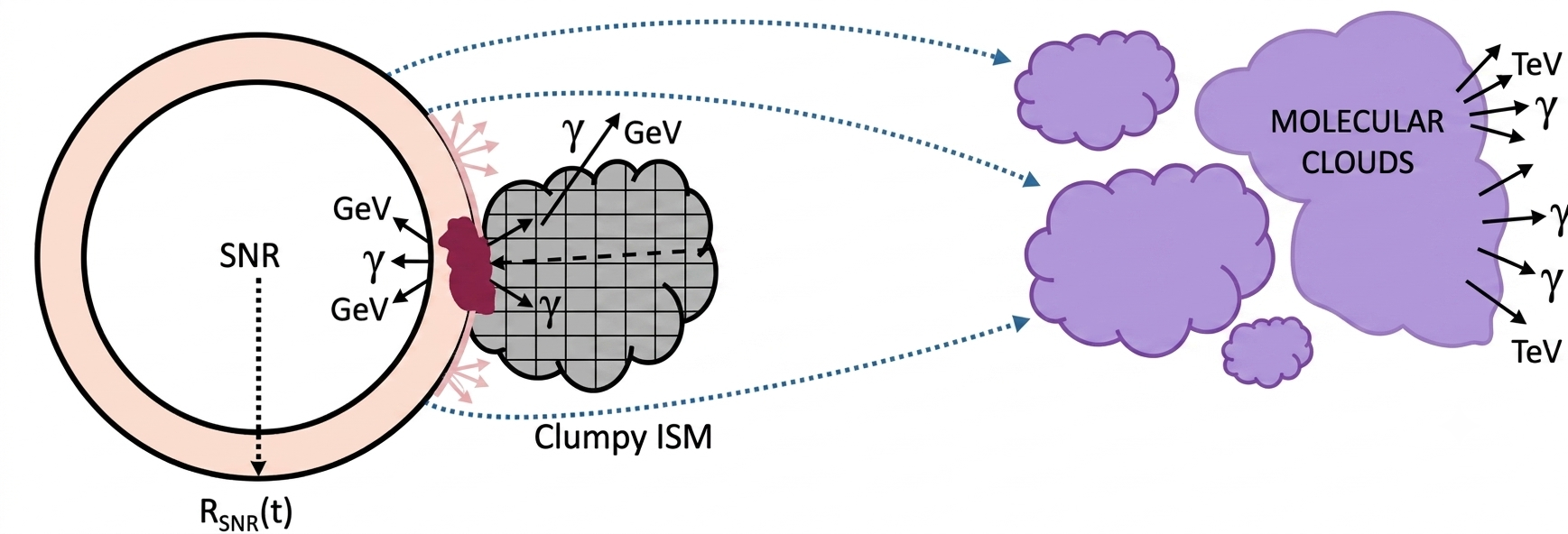}
    \caption{A simple schematic representation of the interaction between the SNR and the GMCs, resulting in the observed multiwavelength emission.}
    \label{fig:SNR_MC_schema}
\end{figure}

In the scenario of a hypothetical SNR located in this region, and in modelling particle transport and molecular-cloud illumination, we adopt the formalism developed in \cite{Gabici_MC_Illumination2009, Mitchell_2021MNRAS.503.3522M, Oka_2025ApJ...989..137O}. This framework is notably different from earlier approaches that assume a spatially homogeneous target cloud. To reproduce the observed GeV $\gamma$-ray spectrum, we additionally account for particles that remain confined within the SNR shell (see Fig.~\ref{fig:SNR_MC_schema}), whose interactions can explain the Fermi-LAT data, in contrast to the synchro-curvature emission scenario proposed by \cite{sarkar24}. Within an angular radius of $20^\circ$, approximately ten SNRs have been identified \citep{Ferrand12, Tsuji25}, some of which may also contribute to the measured emission. However, it is important to emphasize that no SNR has been detected within the field of view of LHAASO.

\begin{figure}
    \centering
    \includegraphics[width=0.95\linewidth]{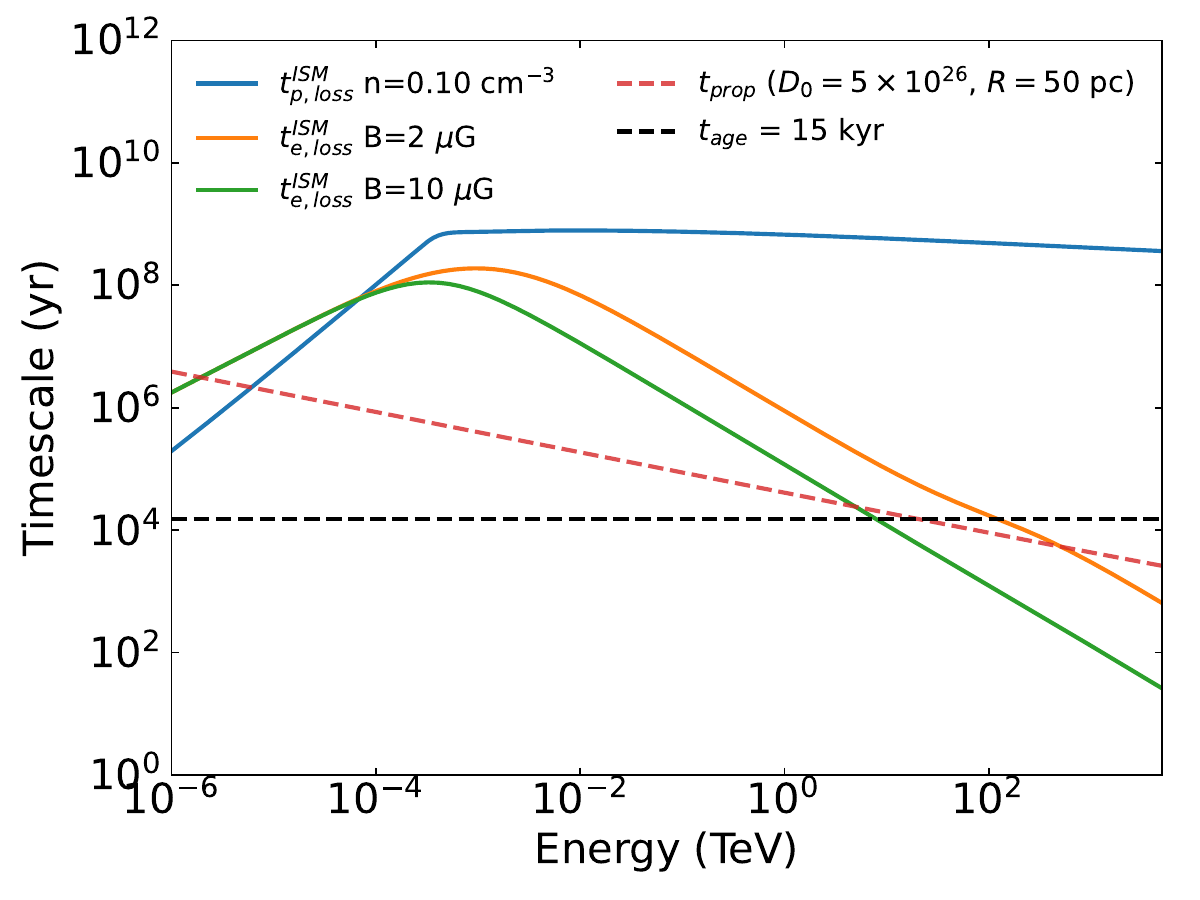}
    \caption{Cooling timescales of protons and electrons propagating through the ISM with gas density $n = 0.1\;\mathrm{cm}^{-3}$ and magnetic field strength $2\;\mathrm{and}\;10\;\mu\mathrm{G}$. The characteristic propagation timescale is shown as the sum of the diffusion and crossing times over a distance of $50\;\mathrm{pc}$, for suppressed diffusion relative to the average Galactic diffusion coefficient.}
    \label{fig:ISM_loss_escape_times}
\end{figure}

Prior to initiating the modelling, we studied the particle transport and energy-loss timescales within GMCs and the ISM. These timescales are predominantly computed following the formalism presented in \citet{Gabici_MC_Illumination2009, Mitchell_2021MNRAS.503.3522M}. The characteristic energy-loss times of protons and electrons in the ISM, together with their diffusion time to reach a distance of 50 pc for a diffusion constant $D_0 = 5\times10^{26}\,\mathrm{cm}^2\,\mathrm{s}^{-1}$, are shown in Fig.~\ref{fig:ISM_loss_escape_times}. The figure indicates that, considering an ambient gas density of $0.1\;\mathrm{cm}^{-3}$, the $p$–$p$ energy losses are negligible and protons can reach the clouds essentially unaffected. Electrons, in contrast, lose a substantial fraction of their energy via synchrotron radiation. If electrons are assumed to escape the system at the same time with protons, then, for a system age of $15\;\mathrm{kyr}$ and a magnetic field strength of $10\;\mu\mathrm{G}$, electrons lose nearly all of their energy before reaching the clouds, and their contribution at the cloud locations can be safely neglected. For a weaker magnetic field of $2\;\mu\mathrm{G}$, some electrons may still survive; however, given an electron-to-proton ratio of $K_{ep}=10^{-3}$ typical for early evolution of SNRs \citep{2021MNRAS.508.6142M, 2012A&A...538A..81M, berezhko2008cosmic,Berezhko:2013ccp}, their leptonic contribution to the observed $\gamma$-ray emission remains negligible and can be ignored.

\begin{figure}
    \centering
    \includegraphics[width=0.95\linewidth]{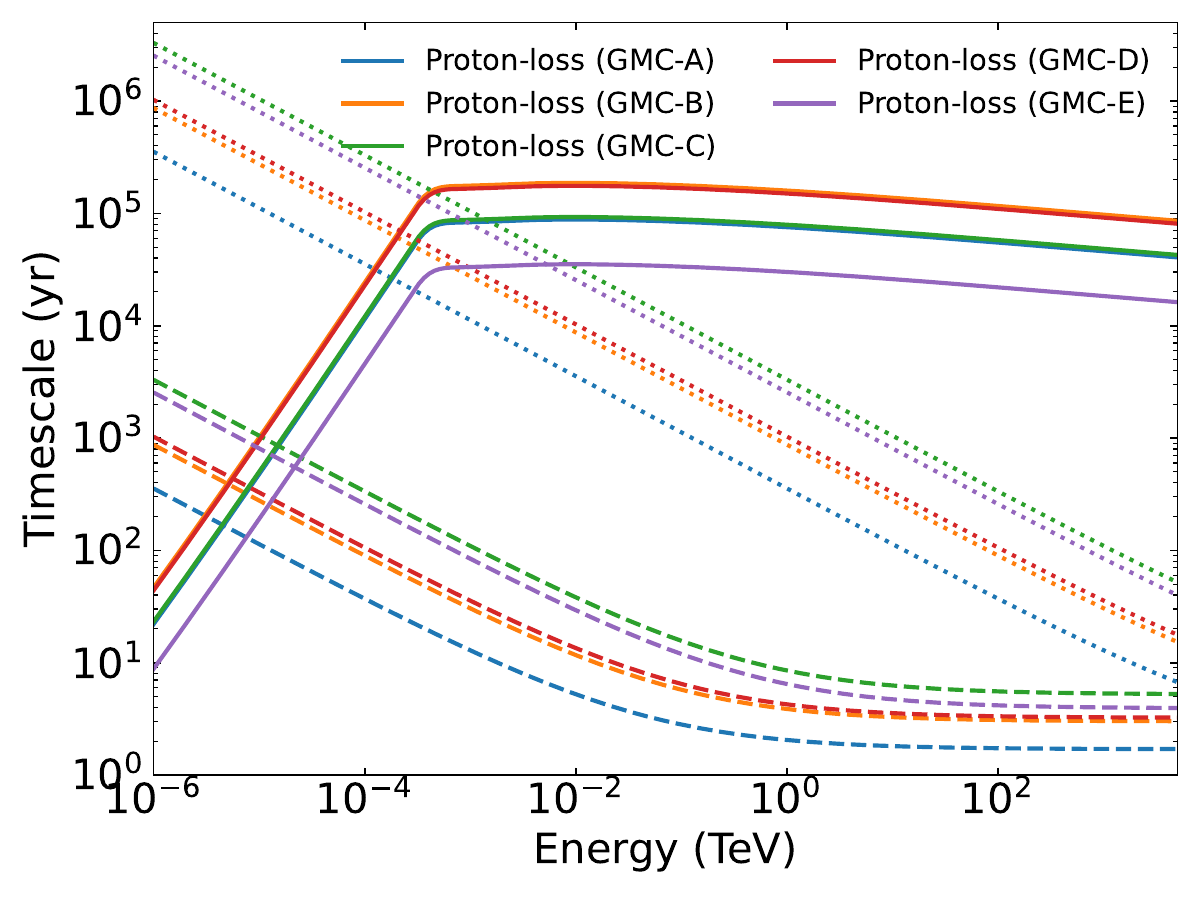}
    \caption{Cooling and propagation timescales of protons compared with the characteristic propagation time, defined as the sum of the diffusion and crossing times within the GMCs, for standard ($\chi = 1$) and suppressed ($\chi = 10^{-3}$) diffusion relative to the average Galactic value.}
    \label{fig:GMC_loss_escape_times1}
\end{figure}

In Fig.~\ref{fig:GMC_loss_escape_times1}, the energy-loss timescales (left panel: protons, right panel: electrons) and the diffusion timescales for both the standard and suppressed diffusion scenarios are shown as solid, dashed, and dotted curves, respectively, for each individual GMC. From this figure, we can see three prominent $p$-$p$ loss curves due to similar gas densities among some GMCs. It is also evident that at low energies, the diffusion timescale for both protons and electrons exceeds their corresponding energy-loss timescales. As a result, low-energy particles are expected to lose nearly all of their energy before escaping from the GMCs. At higher energies, the diffusion timescale becomes significantly shorter than the proton energy-loss timescale, implying that protons can escape the clouds before dissipating a substantial fraction of their energy, and the energy loss will not attenuate the proton spectrum. These results indicate that merely suppressing the diffusion coefficient in the clouds is likely insufficient to prevent low-energy accelerated particles from freely interacting with the clouds and reproducing the LHAASO WCDA observations in the sub-TeV energy range. Instead, an energy-dependent particle escaping from a distant source appears necessary and is expected to play a pivotal role in suppressing the particle population that reaches the clouds within this energy interval. At higher energies, the residence time of particles within the considered GMCs becomes short, allowing UHE particles to escape from the interacting MCs. Consequently, a cutoff at TeV energies is expected to appear in the observed $\gamma$-ray spectrum at its highest detectable energies. This behaviour may also be related to the intrinsic maximum energy attainable at the accelerator (source) itself, which is fixed at 1 PeV in our model.

Under the adopted SNR scenario, we assume that the remnant originated from a supernova explosion that occurred within the last several thousand years and has since expanded into a low-density cavity whose swept-up shell is currently not observable. CRs are accelerated at the shock front via the standard DSA mechanism and subsequently escape from the source region. Here, we consider a momentum-dependent escape of particles, characterized by an escape time \(t_{\mathrm{esc}}(p)\), such that the radius at which particles of energy \(E\) escape is given by
\[
R_{\mathrm{esc}} = R_{\mathrm{SNR}}\bigl(t = t_{\mathrm{esc}}(E)\bigr)
\]
For the time-dependent escape of particles, we adopt the relation
\begin{equation}
    t_{\mathrm{esc}}(p) = t_{\mathrm{sed}}\left(\frac{p}{p_M}\right)^{-1/\beta} \;\; \text{yr},
    \label{equ:esc_time}
\end{equation}
where \(t_{\mathrm{sed}}\) is the onset time of the Sedov–Taylor phase \citep{2021MNRAS.508.6142M, 2019MNRAS.490.4317C}, \(p\) is the particle momentum, \(p_M\) is the maximum momentum, and \(\beta\) is a parameter describing the momentum dependence of the escape process. The temporal evolution of the SNR radius is described by \citep{Mitchell_E_2023MNRAS.520..300M}
\begin{equation}
    R_{\mathrm{SNR}}(t) = 0.31 \left(\frac{E_{\mathrm{SN}}/10^{51}\,\mathrm{erg}}{(n/1\,\mathrm{cm}^{-3})(\mu_1/1.4)}\right)^{1/5} \left(\frac{t}{\mathrm{yr}}\right)^{2/5} \;\; \text{pc},
\end{equation}
where \(E_{\mathrm{SN}}\) is the kinetic energy released in the supernova explosion, \(n\) is the ambient number density, and \(\mu_1\) is the mean molecular weight normalized to 1.4. We assume that the energy spectrum of cosmic-ray protons and electrons that remain confined within the supernova shell, follows a power-law distribution with spectral index $\alpha$ and an exponential cutoff in energy, which can be expressed as \citep{Oka_2025ApJ...989..137O}
\begin{equation}
    J_{p,\mathrm{shell}}(E,t) = N_0\, E^{-\alpha} \exp\left(-\frac{E}{E_{\mathrm{now}}}\right),
\end{equation}
and,
\begin{equation}
    J_{e,\mathrm{shell}}(E,t) = K_{ep}\, N_0\, E^{-\alpha} \exp\left(-\frac{E}{E_{\mathrm{now}}}\right),
\end{equation}

where $E_{\mathrm{now}}$ represents the current maximum energy can be calculated following Eq.~\ref{equ:esc_time} \citep{Oka_2025ApJ...989..137O,2019MNRAS.490.4317C}, and $K_{ep}$ denotes the electron-to-proton normalization ratio. The normalisation constant is denoted as $N_0$, which is determined from the total energy released by the SNR and subsequently converted into the energy budget of the accelerated CRs. It can be expressed as

\begin{equation}
    N_0 = 
\begin{cases} 
\dfrac{\eta E_{\mathrm{CR}}}{\ln(E_{\max}) - \ln(E_{\min})} & \text{for } \alpha = 2, \\[15pt]
\dfrac{\eta E_{\mathrm{CR}} (2 - \alpha)}{E_{\max}^{2-\alpha} - E_{\min}^{2-\alpha}} & \text{otherwise.}
\end{cases}
\label{equ:normSNR}
\end{equation}
where $\eta$ denotes the acceleration efficiency of the SNR, which we fix to 30\% \citep{Gabici_MC_Illumination2009}. The quantity $E_{\mathrm{CR}}$ represents the total energy transferred to CRs. The minimum and maximum energies of the particles accelerated during the Sedov phase are indicated by $E_{\min}$ and $E_{\max}$, and are set to 280 MeV and 1 PeV, respectively.


The spatial and energetic distribution of particles that have escaped from the SNR with energies $E > E_{\mathrm{esc}}$ and are located at distances $R > R_{\mathrm{esc}}$ at the current evolutionary time $t = t_{\mathrm{age}}$ can be written as \citep{Mitchell_2021MNRAS.503.3522M}

\begin{equation}
    J_p(E,R,t) = N_0\,E^{-\alpha} \exp\left(-\frac{E}{E_{p,\mathrm{cut}}}\right) f(E,R,t),
\end{equation}

where the cutoff energy $E_{p,\mathrm{cut}}$, on the other hand, could be associated with the escape of particles from the interaction region or the maximum energy achievable by the SNR during its Sedov phase. Due to this ambiguity, we leave this as a free parameter in our model fitting. The function $f(E,R,t)$ represents the probability density function of particles that have escaped the SNR, propagated diffusively through the ISM, and eventually reached the MCs, where they interact and produce TeV $\gamma$-ray emission, as written in Eq.~\ref{eq:pdfSNR} \citep{Mitchell_2021MNRAS.503.3522M},

\begin{equation}
    f(E,R^\prime,t^\prime) \approx \frac{f_0}{\pi^{3/2} R_d^3}
    \exp\left[-\frac{(\alpha - 1) t^\prime}{\tau_{pp}} - \frac{R^{\prime 2}}{R_d^2}\right],
    \quad \text{for } t > t_{\mathrm{esc}}(E),
    \label{eq:pdfSNR}
\end{equation}

where $R^\prime$ denotes the effective path length available for particles to travel through the ISM before intercepting the MC, after having escaped from the SNR. If $d$ is the centre-to-centre separation between the SNR and the cloud, then the available distance is $R^\prime = d - R_{\rm esc} - R_{\rm cloud}$, where $R_{\rm esc}$ is the radius of the SNR at the time of particle escape and $R_{\rm cloud}$ is the characteristic radius of the cloud. Likewise, $t^\prime$ represents the effective propagation time of particles in the ISM before reaching the cloud and is given by $t^\prime = \min\bigl(t_{\rm age} - t_{\rm esc},\, \tau_{\rm ISM}\bigr)$ where \(t_{\rm age}\) is the age of the SNR, \(t_{\rm esc}\) is the escape time of the particles from the shock, and $\tau_{\rm ISM}$ is the diffusion time in the ISM.

The diffusion length scale and the characteristic timescale for inelastic $p$-$p$ interactions are denoted by \(R_d\) and \(\tau_{pp}\), respectively \citep{Mitchell_E_2023MNRAS.520..300M}. The diffusion radius \(R_d\) is defined as 
\begin{equation}
    R_d(E,t^\prime) = 2 \sqrt{D(E)\, t^\prime \, \frac{\exp\!\bigl(t^\prime \delta / \tau_{pp}\bigr) - 1}{t^\prime \delta / \tau_{pp}}} \;\; \text{cm},
\end{equation}
where \(D(E)\) is the energy-dependent diffusion coefficient, and \(\delta\) is the diffusion index. The diffusion coefficient is considered here as power law with energy as:
\begin{equation}
    D(E) = \chi D_0\left( \frac{E}{10\; GeV}\right)^\delta  \;\; \text{cm$^2$ s$^{-1}$}
\end{equation}
where the suppression factor $\chi$ relates to the level of turbulence in the propagation region, and the reference diffusion coefficient $D_0=10^{28}$ cm$^2$ s$^{-1}$ is evaluated at 10 GeV.


\begin{table*}
    \centering
    \renewcommand{\arraystretch}{1.25}
    \setlength{\tabcolsep}{4pt}
    \caption{Model parameters used for the SNR and surrounding interstellar medium. The first block lists fitted parameters with their 1$\sigma$ uncertainties and allowed fitting ranges; the second block lists fixed (adopted) parameters.}
    \label{tab:snr_model_parameters}
    \begin{tabular}{llccc}
        \hline\hline
        Model parameters & Symbol [unit] & MCMC Fit values  & $\chi^2_{\rm min}$ & Prior / range \\ \hline
        Age of the SNR & $t_{\mathrm{SNR}}$ [kyr] & $49.4^{+22.3}_{-15.0}$ & 38.7 & $0.5$--$100$ \\
        Slope of diffusion coefficient & $\delta$ & $0.37^{+0.06}_{-0.05}$ & 0.41 & $0.3$--$0.6$ \\
        ISM diffusion suppression factor & $\chi$ & $2.26^{+6.18}_{-0.97}\times10^{-3}$ & $1.2\times10^{-3}$ & $10^{-3}$--1 \\
        Slope of momentum distribution & $\beta$ & $2.80^{+0.13}_{-0.19}$ & 2.81 & 2--3 \\
        Spectral index of CRs & $\alpha$ & $2.13^{+0.04}_{-0.05}$ & 2.18 & 1.8--2.4\\ 
        Sedov time & $t_{\mathrm{sed}}$ [yr] & $302.5^{+131.6}_{-76.0}$ & 255 & 200--2000 \\
        Gas density near the SNR & $n_{\mathrm{shell}}$ [cm$^{-3}$] & $36.77^{+2.41}_{-3.91}$ & 39.7 & 1--40 \\
        TeV cut-off energy & $E_{\mathrm{p,cut}}$ [TeV] &  $839.79^{+116.77}_{-174.32}$ & 739 & 1--1000 \\
        Distance to GMC-A & $d_1$ [pc] & $70.57^{+48.58}_{-30.83}$ & 74.8 & 30--150 \\
        Distance to GMC-B & $d_2$ [pc] & $50.66^{+63.88}_{-15.13}$ & 33.4 & 30--150 \\
        Distance to GMC-C & $d_3$ [pc] & $38.82^{+12.69}_{-6.30}$ & 31.7 & 30--150 \\
        Distance to GMC-D & $d_4$ [pc] & $88.75^{+44.19}_{-44.78}$ & 71.8 & 30--150 \\ \hline
        SNR initial energy & $E_{\mathrm{SN}}$ [erg] & \multicolumn{2}{c}{$10^{51}$} & fixed \\
        Acceleration efficiency & $\eta$ & \multicolumn{2}{c}{30\%} & fixed \\
        Electron-to-proton ratio & $K_{\mathrm{ep}}$ & \multicolumn{2}{c}{$10^{-3}$} & fixed \\\hline
    \end{tabular}
\end{table*}

The normalization factor $ f_0 $ is introduced to account for the injection of particles at a time-dependent radius $ R_{\text{SNR}}(t) $ from a point-like origin (e.g. the centre of the SNR) \citep{Aharonian_1996A&A...309..917A}. This factor is obtained by imposing the normalization condition $ \int f(E,R,t)\, dV \;=\; \int f(E,R,t)\, 4\pi R^2\, dR \;=\; 1,$ which leads to the solution \citep{Mitchell_E_2023MNRAS.520..300M}
\begin{equation}
    f_0 \;=\; \frac{\sqrt{\pi}\, R_d^3}{\bigl(\sqrt{\pi} R_d^2 + 2\sqrt{\pi} R_{\text{SNR}}^2\bigr) R_d + 4 R_{\text{SNR}} R_d^2}
\end{equation}


Considering the model described above, we fit the observational data using the Markov Chain Monte Carlo (MCMC) sampler emcee \citep{emcee}. Within this framework, we assume that the GeV emission detected by Fermi-LAT originates from a localized region of the SNR shell interacting with the surrounding ambient gas, whereas the TeV $\gamma$-ray emission detected by LHAASO is produced in MCs distributed radially around the SNR at different distances, as projected along the line of sight within the detector’s field of view.

\begin{figure}
    \centering
    \includegraphics[width=0.9\linewidth]{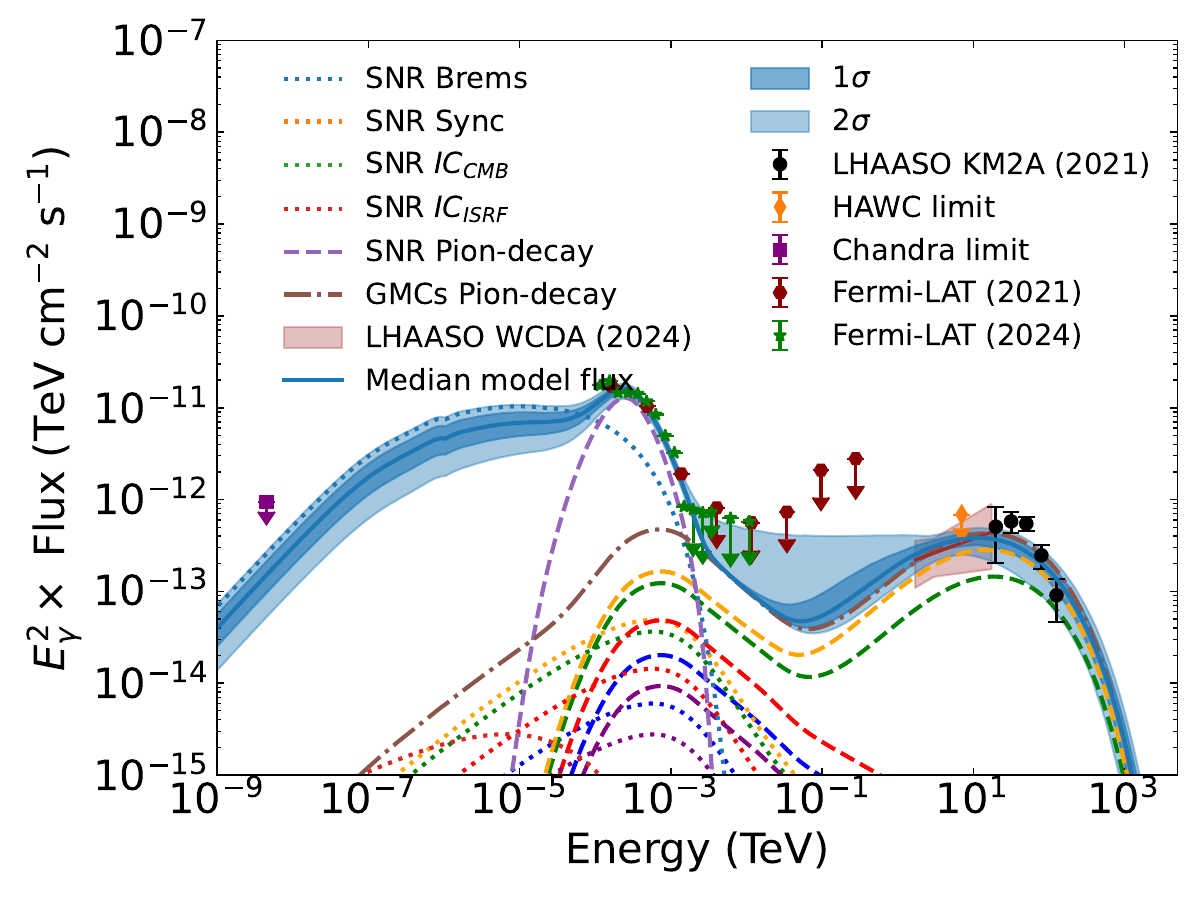}
    \caption{The multiwavelength modelling of the LHAASO source is presented under the assumption of a past SNR explosion as the origin of the $\gamma$-ray emissions. In the figure, the dashed curve represents the emission component arising from hadronic interactions, specifically $\gamma$-rays produced via neutral pion decay, while the dotted curve denotes the contribution from leptonic processes corresponding to the $\chi^2_{min}$. The $1\sigma\;\mathrm{and}\;2\sigma$ containment band is shown by the light and dark blue shaded region, where the solid line corresponds to the median emission curve of the model considered here. See Tab. \ref{tab:snr_model_parameters} for model parameters.}
    \label{fig:snr_mc_fit}
\end{figure}

The free and fixed parameters of our model are summarized in Tab.~\ref{tab:snr_model_parameters}, where we list the fit values, their corresponding $1\sigma$ uncertainties, and the maximum-likelihood parameters associated with the minimum $\chi^2$ value, together with the prior ranges adopted in the MCMC analysis. Using this model, we perform a simultaneous fit to the observed multiwavelength data from the GeV to the TeV band, as shown in Fig.~\ref{fig:snr_mc_fit}. The $1\;\mathrm{and}\;2\sigma$ containment interval of the posterior predictive distribution is also displayed in the same figure, along with the individual contributions of the different emission components corresponding to the best-fit parameter set. The best-fit model yields a Sedov timescale of approximately 255 years, which is consistent with the relatively short Sedov phase expected for a Type Ia supernova remnant. At the present age of 38.7 kyr, the SNR shell has expanded to a radius of 10.16 pc, with a portion of the shell interacting with clumpy ISM of density $\sim 39.7\;\mathrm{cm^{-3}}$, producing the GeV $\gamma$ rays detected by Fermi-LAT. This ambient density is relatively low compared to that of the GMCs identified in the vicinity. In this work, we assume a suppression of the diffusion coefficient relative to the canonical Galactic value, attributable to enhanced magnetic turbulence generated in the vicinity of the source by the escaping CRs themselves \citep{2011MNRAS.415.3434F,2013ApJ...768...73M,2018MNRAS.474.1944D}, similar to that around pulsar wind nebulae \citep{2017Sci...358..911A}. The fit favours a suppression factor of $\sim 10^{-3}$, corresponding to a reference diffusion coefficient of $D_0 \simeq 10^{25}\,\mathrm{cm^{2}\,s^{-1}}$ with a diffusion index $\delta = 0.41$, intermediate between the values expected for Kolmogorov- and Kraichnan-type turbulence. For a fixed $\delta$, an increase in $D_0$ shifts the TeV $\gamma$-ray spectral peak to lower energies, whereas for larger values of $\delta$ the preferred $D_0$ becomes smaller \citep{2022PASJ...74..625O, Oka_2025ApJ...989..137O}. In this modelling, the spectral index of CRs confined near the SNR shell is assumed to be similar to that of particles that have escaped the source region, with a best-fit value of $\sim 2.18$. However, \citet{2010A&A...513A..17O} argue that the spectral index in the vicinity of the SNR shell can differ from that of the escaping particles, and propagation models indicate that the spectral index of CRs escaping typical Galactic sources is expected to lie in the range $2.2$–$2.4$ \citep{PhysRevD.95.083007,2012ApJ...752...69O}. Given the current lack of detailed information about the source, future observations that better constrain the intrinsic properties of the SNR and its environment will enable more precise modelling and parameter estimation, but such an investigation is beyond the scope of this work.

\section{Neutrinos from LHAASO J0341+5258}
\label{Sec:neutrinos}
\begin{figure}
    \centering
    \includegraphics[width=0.9\linewidth]{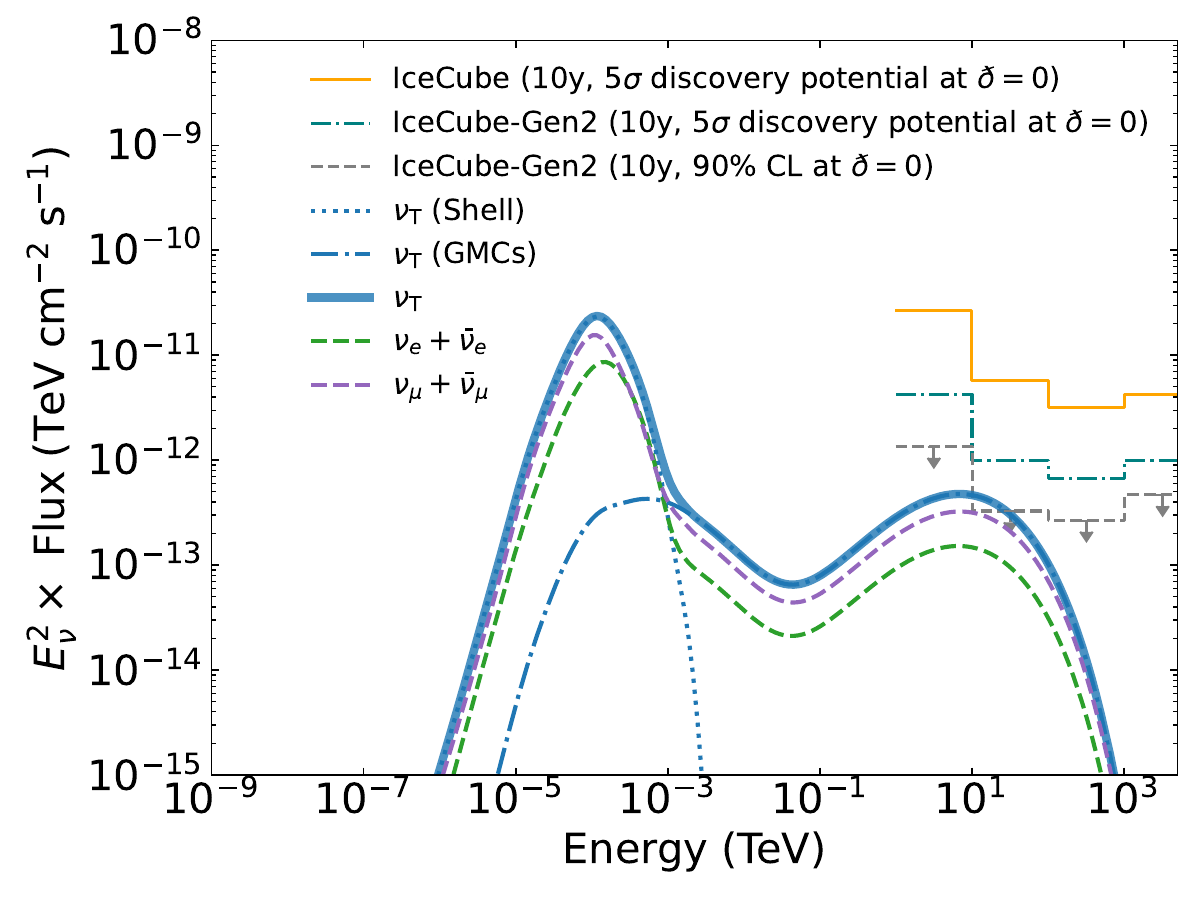}
    \caption{The electron, muonic and total neutrino flux from the interaction of cosmic rays near the source region is illustrated here along with the 10-year IceCube sensitivity curves.}
    \label{fig:snrNu}
\end{figure}

We have also evaluated the total neutrino emission from the region under the assumption of a hadronic scenario, in which the observed high-energy $\gamma$-ray emission is produced via inelastic $p$–$p$ interactions and is therefore accompanied by neutrino production. To compute the neutrino yield from $p$–$p$ interactions, we adopt the differential production cross section of \citet{2021PhRvD.104l3027K} for primary proton energies $E_p \geq 4\,\mathrm{GeV}$. For $E_p < 4\,\mathrm{GeV}$, where the parameterization of \citet{2021PhRvD.104l3027K} is not optimized, we instead use the prescription of \citet{2006ApJ...647..692K} as implemented in the AAfrag package\footnote{\url{https://github.com/aafragpy/aafragpy}}. In both regimes, the neutrino production rate is obtained by convolving the assumed proton distribution and the ambient gas density with the corresponding energy-dependent $p$–$p$ cross sections. The resulting all-flavor neutrino spectrum corresponding to the minimum-$\chi^2$ fit to the $\gamma$-ray data is shown in Fig.~\ref{fig:snrNu}. The separate $\nu_e$ and $\nu_\mu$ components are also displayed, along with the relevant IceCube sensitivity curves. The predicted neutrino flux from the source remains below the current IceCube detection threshold. For IceCube-Gen2, even with an exposure of 10 years, the expected signal would only reach the 90\% confidence level at declination $\eth = 0^\circ$, with the sensitivity degrading toward higher declinations.

\section{Conclusions}
\label{Sec:conclusion}
The advent of UHE $\gamma$-ray astronomy has transformed our understanding of Galactic particle accelerators. Recent observations have revealed an increasingly diverse population of candidate PeVatrons, including SNRs, PWNe, massive star-forming regions, stellar clusters, and MQs, highlighting that particle acceleration to PeV energies may occur in a wider range of environments than previously anticipated \citep{lhaaso24}. At the same time, a growing number of UNID and dark UHE sources present a new challenge. Unlike established PeVatron candidates, these objects lack a clear accelerator counterpart within their emission regions. Among the sources reported by LHAASO \citep{lhaaso21,lhaaso24}, LHAASO J0341+5258 stands out as a particularly compelling case. The absence of any obvious powerful accelerator, together with the strong spatial correlation between the observed $\gamma$-ray emission and nearby MCs, motivates an alternative perspective in which the surrounding medium plays the central role. This raises a fundamental question: can MCs preserve the signatures of past PeV particle acceleration by acting as long-lived reservoirs of CRs? Addressing this question is crucial not only for understanding the origin of LHAASO J0341+5258, but also for interpreting the growing population of UNID UHE emitters associated with dense molecular environments. In this context, MCs may serve as unique probes of both particle transport and past accelerator activity, offering an alternative pathway to uncover the nature of hidden Galactic PeVatrons.

In this paper, the observed emission from the source J0341+5258, is modeled using two independent scenarios: (i) a time-dependent, source-independent hadronic model implemented with \textsc{GAMERA}, where particles are injected into the ISM by an unknown accelerator, where they propagate and undergo energy losses, escape and interact with the molecular gas observed in the region, and (ii) an analytical framework describing the interaction between SNR and MCs, where the GeV-TeV emisison is mapped by the SNR itself and the particles escaping from the SNR illuminating the MC. 

In the source-independent model, we follow the evolution of the accelerated particle population and explore whether their subsequent propagation and interaction with the surrounding environment can reproduce the observed $\gamma$-ray spectrum, without focusing on the nature of the accelerator itself. We considered a population of particles injected into the ISM and propagating over a distance of 50 pc. During propagation of, the particles undergo both spatial diffusion and energy losses. We derive the particle population escaping from the ISM following the methodology outlined in Sec.~\ref{subsec:ISM_prop} and subsequently inject the resulting spectrum into the MC. The target cloud was modeled as a single effective MC whose physical properties were obtained by averaging all MCs identified by \citet{Tsuji25} within the $\sim1^\circ$ region encompassing the full LHAASO emission extent at the 68\% containment radius (see Table~\ref{tab:MC_tab}). The injected particles were then evolved within this averaged cloud for 100 yrs and the resulting spectral energy distribution, also considering nuclear enhancement, is shown in Fig.~\ref{fig:SED_illum}. The resulting spectrum supports a CR illumination scenario in which particles accelerated by a powerful source escape, propagate through the ISM, and eventually interact with MCs where they remain efficiently confined over extended timescales Within this framework, the UHE emission detected by LHAASO is naturally reproduced through hadronic $p$-$p$  interactions. However, Fig.~\ref{fig:SED_illum} also highlights a limitation of the model: while the UHE component is successfully reproduced, the observed GeV emission is not. This indicates the presence of an additional low-energy particle population. The total particle energetics inferred from the illumination model point toward an accelerator operating during an efficient acceleration phase, possibly a SNR. Motivated by this result, we subsequently explore a distant SNR scenario in which the GeV emission originates from the remnant itself, while the UHE component is produced by particles that have escaped the accelerator and illuminated the surrounding MC.

Our analytical modelling supports a scenario in which a relatively older, likely Type-Ia SNR, based on its relatively short Sedov time \citep{Mitchell_2021MNRAS.503.3522M}, is interacting with MCs, giving rise to the observed GeV-TeV emission through hadronic processes under conditions of locally suppressed diffusion. The inferred turbulence-driven reduction of the diffusion coefficient and the relatively hard CR spectrum point to efficient particle acceleration and confinement in the vicinity of the remnant. While these results provide a self-consistent description of the multiwavelength data, more detailed 3D observations of both the SNR structure and its ambient medium are required to refine the physical picture and to test the assumptions underlying the present model.

In conclusion, we show that the source-independent illumination scenario can explain the UHE emission seen by LHAASO. At the same time, in the absence of any confirmed PWN candidate, we consider an evolved SNR as a plausible origin of the observed GeV/TeV emission. However, a PWN or a composite SNR+PWN system could also reproduce a similar spectral energy distribution. Future multiwavelength and higher-sensitivity observations will be required to distinguish between these scenarios and to achieve a more comprehensive understanding of the emission from this region.

\begin{acknowledgements}
The authors thank Naomi Tsuji for generously providing the CO intensity map presented in Fig 1. We also acknowledge Cong Li for providing the LHAASO data. AR and JCJ acknowledge the support and hospitality of INAF-IAPS, Rome, during their visit, where a major part of this work was completed. Their visit was funded through the INAF Mini-Grant PERLA (PI: MC). AS acknowledges the financial support by PNRR - CTA+ PROGRAM (Proposal: IR0000012) PhD fellowship, funded by the European Union - NextGenerationEU and approved by the MUR.
\end{acknowledgements}

%
\bibliographystyle{bibtex/aa} 
\bibliography{Ref} 
\begin{appendix}




\onecolumn
\section{Proton particle population pre- and post-ISM propagation} 
\label{App:Proton_part_story}

\begin{figure}[!ht]
    \centering
    \includegraphics[height=0.35\linewidth]{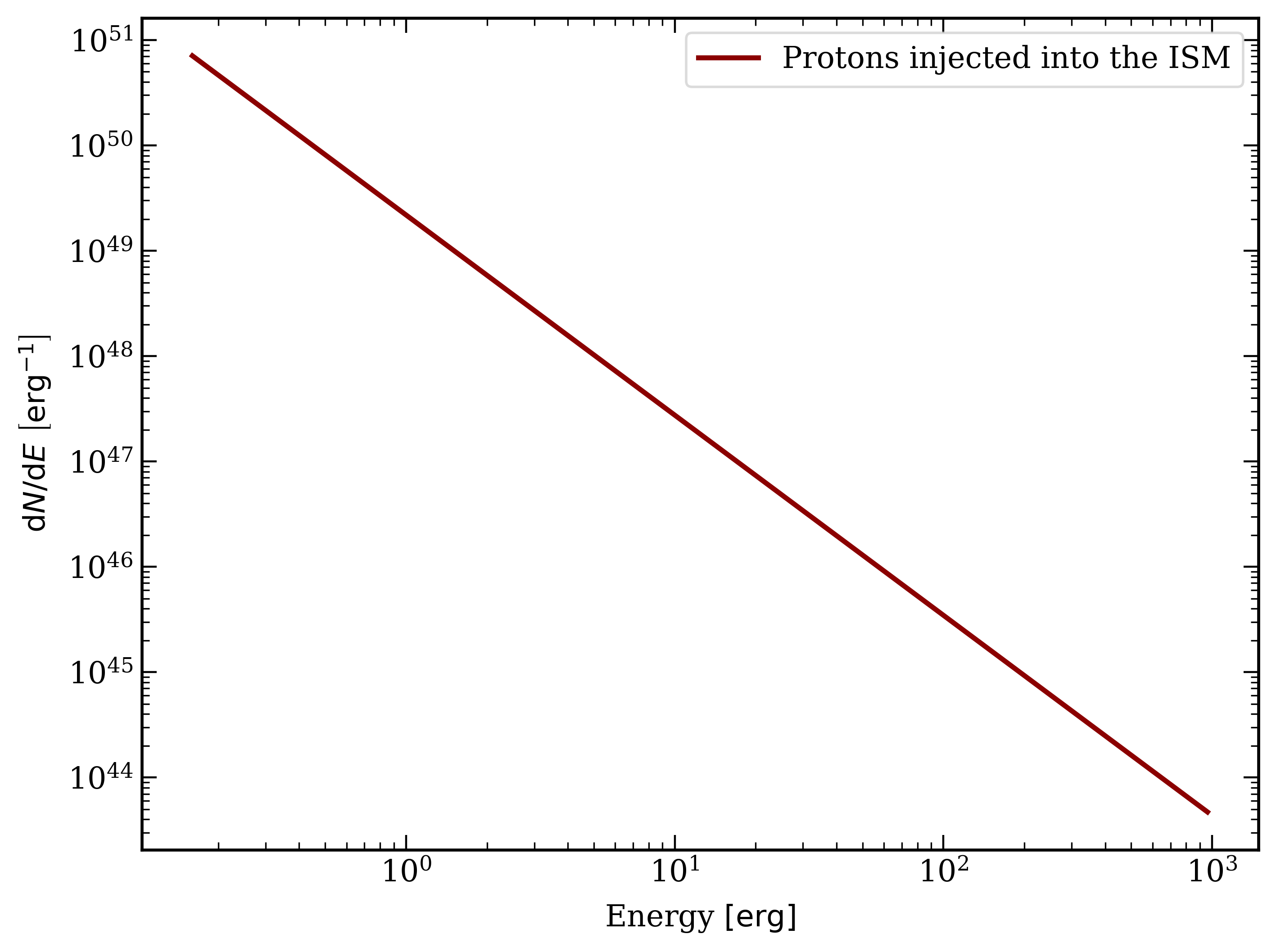}
    \includegraphics[height=0.35\linewidth]{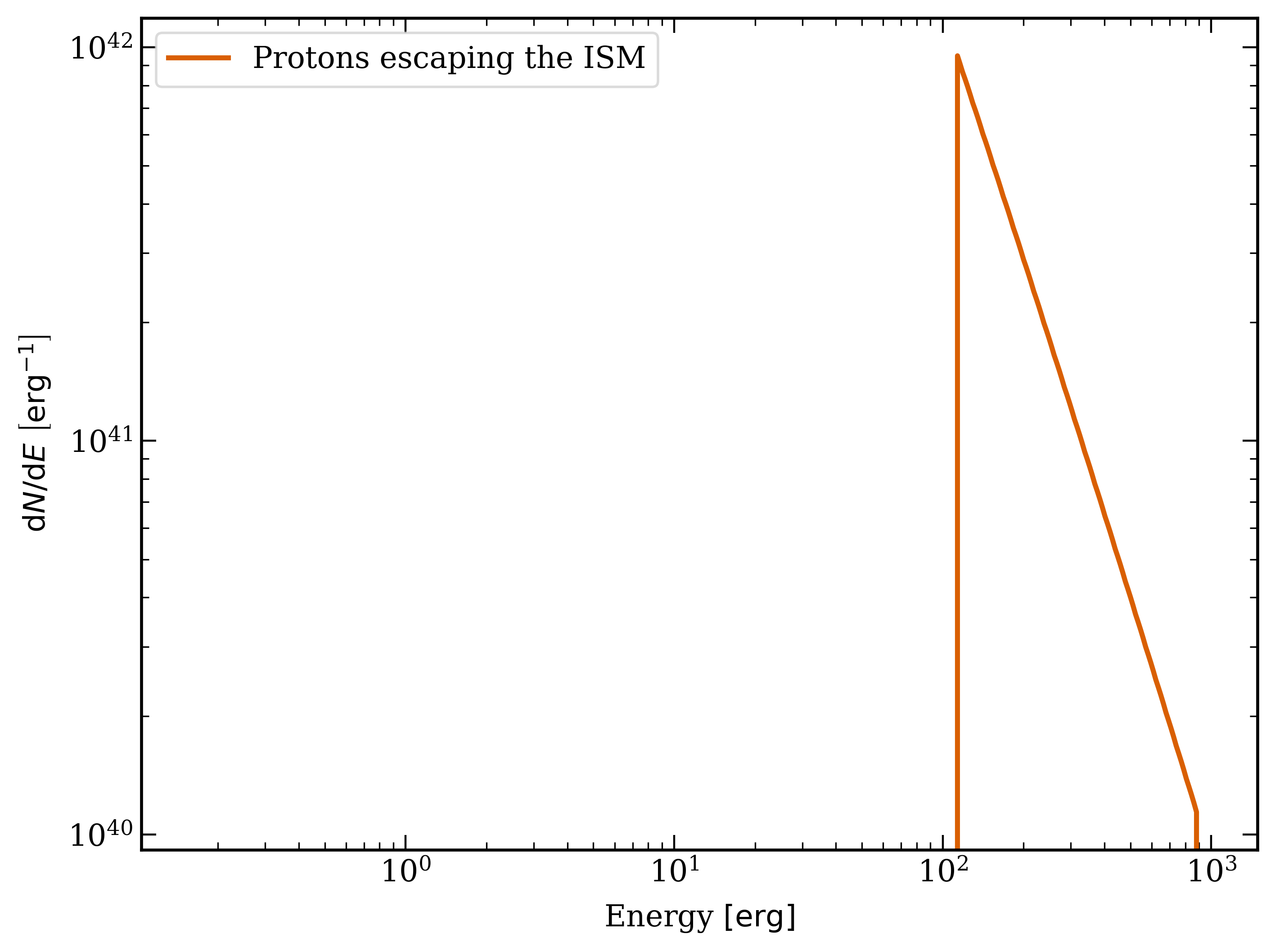}
    \caption{The proton particle population injected into the ISM (\textit{left panel}) and the one escaping the ISM (\textit{right panel}).}
    \label{fig:Pre_Post_ISM_protons}
\end{figure}

Here we present the evolution of the proton population as particles are injected into the ISM. Following the formalism described in Sec.~\ref{subsec:ISM_prop}, we account for diffusion and energy losses, and extract the particles escaping the ISM under the horizon condition $t_{D} <\,t_{prop}$. The resulting escaped particle spectrum is subsequently injected into the MC environment to derive the final spectral energy distribution shown in Fig.~\ref{fig:SED_illum}.

\section{MCMC Fitting Parameters} 
\label{App:fitc}
\begin{figure}
    \centering
    \includegraphics[width=0.6\linewidth]{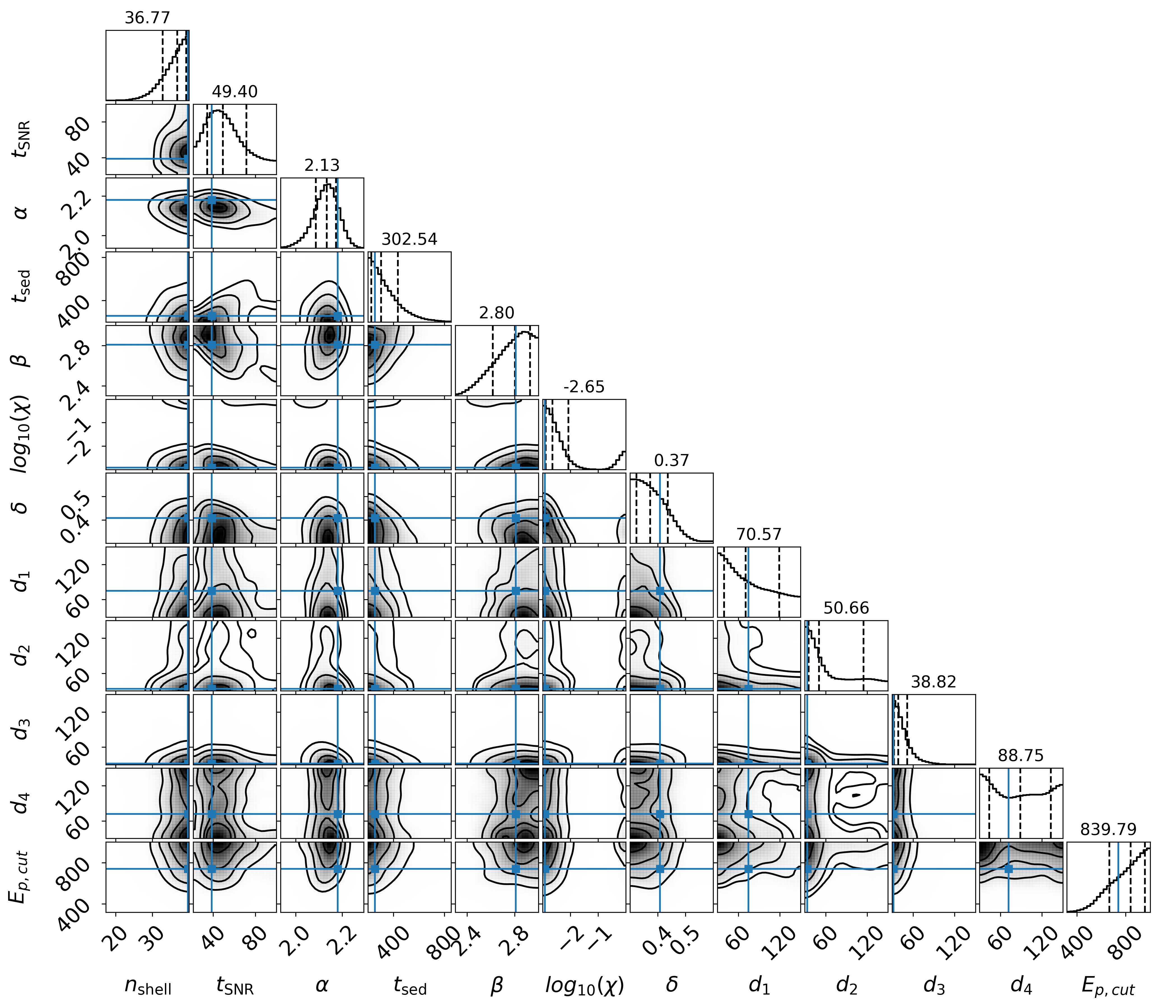}
    \caption{The correlations among the various parameters included in the fit are represented in a corner plot, where the dashed lines indicate the 68\% credible (containment) interval for each parameter, and the blue line denotes the corresponding best-fit parameter.}
    \label{fig:cornor}
\end{figure}

To infer the optimal parameter values, we executed 10,000 sampling iterations of the model employing 200 walkers. The initial 5,000 iterations were discarded as burn-in to ensure convergence of the Markov chains. The posterior parameter correlations obtained after the burn-in phase are displayed in Fig.~\ref{fig:cornor} as a corner plot.


\end{appendix}
\end{document}